\documentclass[11pt]{cernrep}
\usepackage{graphicx,epsfig,epsf}
\usepackage{here}
\usepackage{amsmath}

\def\PT{\mbox{\scriptsize{PT}}}
\def\NP{\mbox{\scriptsize{NP}}}
\def\UV{{\mbox{\scriptsize UV}}}

\def\LQCD{\Lambda_{\mbox{\scriptsize QCD}}}

\def \alpi {{\alpha_s \over \pi}}
\def\I{{\rm i}}

\def\ford#1#2{
\begin{flushright}
\begin{minipage}{#1\textwidth}
#2
\end{minipage}
\end{flushright}
}

\def\beeq{\begin{eqnarray}}
\def\eeeq{\end{eqnarray}}
\def\beq{\begin{equation}}
\def\eeq{\end{equation}}
\def\d{\widehat{{\rm\bf d}}}
\def\epem{e^+e^-}

\def\la{\mathrel{\mathpalette\fun <}}
\def\ga{\mathrel{\mathpalette\fun >}}
\def\fun#1#2{\lower3.6pt\vbox{\baselineskip0pt\lineskip.9pt
  \ialign{$\mathsurround=0pt#1\hfil##\hfil$\crcr#2\crcr\sim\crcr}}}

\def\tform{t_{\mbox{\scriptsize form}}}
\def\as{\alpha_{\mbox{\scriptsize s}}}
\def\acr{\alpha_{\mbox{\scriptsize crit}}}

\def\lrang#1{\left\langle#1\right\rangle}
\def\cO#1{{\cal O}\left(#1\right)}

\def\sigmatot{\sigma_{\mbox{\scriptsize tot}}}

\def\1{$^{\mbox{\scriptsize\{1\}}}$}
\def\2{$^{\mbox{\scriptsize\{2\}}}$}
\def\qq{q\bar{q}}
\def\al{\alpha}
\def\cb{\beta}

\def\eV{{\rm e\kern-0.12em V}}            \def\MeV{{\rm M}\eV}
 \def\GeV{{\rm G}\eV} 
\def\half{{\textstyle {1\over2}}}

\def\np#1#2#3{{ Nucl.\ Phys.}~{\bf B #1} (#3) #2}
\def\pl#1#2#3{{ Phys.\ Lett.}~{\bf #1 B} (#3) #2}
\def\pr#1#2#3{{ Phys.\ Rev.}~{\bf #1} (#3) #2}
\def\prep#1#2#3{{ Phys.\ Rep.}~{\bf #1} (#3) #2}
\def\prl#1#2#3{{ Phys.\ Rev.\ Lett.}~{\bf#1} (#3) #2}

\def\sjnp#1#2#3{{ Sov.\ J.\ Nucl.\ Phys.}~{\bf #1} (#3) #2}

\def\zp#1#2#3{{ Zeit.\ Phys.}~{\bf C #1} (#3) #2}
\def\ijmp#1#2#3{{ Int.\ J.\ Mod.\ Phys.}~{\bf #1} (#3) #2}
\def\epj#1#2#3{{ Eur.\ Phys.\ J.}~{\bf #1} (#3) #2}

\begin{document}
 \title{QCD Phenomenology\\
{\em Lectures at the CERN--Dubna School, Pylos, August 2002}}
\author{Yu.L. Dokshitzer}
\institute{LPTHE, University Paris-6, France\\ and \\
PNPI, St.~Petersburg, Russia}
\maketitle
 \begin{abstract}
   The status of QCD phenomena and open problems are reviewed
 \end{abstract}

\section*{Foreword}

 The four lectures on ``QCD Phenomenology'' at the $\Sigma{\Large
 \epsilon\rho\nu}$--$\Delta o \upsilon\mu\nu\alpha$ school, delivered in
 the dungeon hall of the magnificent Pylos castle were, naturally,
 labelled by Greek letters and dealt with
\begin{itemize}
\item[$\alpha$ --] the basics of QCD and its main problems,

\item[$\beta$ --] the running coupling and anatomy of the Asymptotic
Freedom,

 \item[$\gamma$ --] QCD partons and the r\^ole of colour in multiple
hadroproduction and

\item[$\delta$ --] non-perturbative corrections to QCD observables. 
\end{itemize}
 In the written version of the lectures I have chosen to concentrate
 on the {\em qualitative}\/ discussion of selected QCD phenomena
 rather than teach you the basic perturbative QCD
 techniques\footnote{A systematic introduction into the physics of
 colour, gluon radiation, parton multiplication etc.\ can be found in
 the Proceedings of another CERN--Dubna school~\cite{Dubna95}.}.
 The selection criterion was as follows. I have picked the topics that
 I find puzzling and/or whose importance I feel have not attracted as
 much attention as they rightfully deserved.

\section{INTRODUCTION. ON GUESSWORKS.}

 \begin{flushright} {\em The spirits were kind enough to say: \\ ``We
 have no respect of cullours''} ~\cite{char1} \end{flushright}

In the late 1970s one could say ``QED was 30 years {\em old}\/''.  In
2003 we cannot but state that ``QCD is 30 years {\em young}\/''.
Dating great discoveries is a delicate business, though.
``The spirit'' in the above Charlotte Fell-Smith's narrative
actually was 
Uriel\footnote{Aleph, vau, resh, yod, aleph, lamed -- ``Fire of God'',
the middle pillar of the Tree of Life and supervisor of Nature
Spirits.}, the one of four Archangels
responsible for fundamental science (and physics in
particular)~\cite{Uriel}. Thus the idea of {\bf colour invariance}, as
communicated by Arch.\  Uriel to Dr.\ John Dee, may be dated back
420 years\footnote{An apparent contradiction with another angelic
message,
``{\em Note the forme of the thing seen. Note the
  cullour}\/''~\cite{chm},
is resolved by accepting that Dr.\ Dee misunderstood the angel: Uriel
meant ``{\em {\bf\em Not} the colour}.''}!  
In 1582 Dr.\ Dee pens down Uriel's detailed instructions on arranging
the conjuring table:
\begin{quote}
  ``{\em The sylk must be of diuers cullors: the most changeable that
  can be got}\/\{ten\}''~\cite{chm}. 
\end{quote}
The idea of colour symmetry could not be put forward more clearly:
``{\em The most changeable, diuers {\em [diverse]} cullors}'' to which
quantum number ``{\em we {\em [angels]} have no respect.}''  However
Dee here adds a confusing marginal remark,
\begin{quote}
 ``{\em The cullor was shewed red and greene interchangeably,}'' \hfill
{\small  Nouemb.\  21.\ Ao 1582}~\cite{chm}.
\end{quote}
This 
suggests 
it may have been $SU(2)$ rather than $SU(3)
$ that Uriel was trying to deliver (unless Dee was colour blind to the
blue part of the spectrum, of which we have no documented evidence).
Tiny details aside, the key idea of the local non-Abelian symmetry had
been clearly present in the angelic message ({\em no respect of
cullours}\/).

Now let us leave Dr.\ John Dee for the time being and stress that the
physics of hadrons always was, and still is, providing puzzles and
inspiration. If 30--40 years ago quantum field theory (QFT) had been
kept in higher respect (which it was not), some general
phenomenological features of hadron interactions that were known then
could have already hinted at QCD as an underlying microscopic theory
of hadrons.

\subsection{Hints from the past}
\vspace{5mm}

\begin{itemize}

\item 
 The fact that in high energy hadron interaction processes {\em
 inelastic}\/ breakup typically dominates over elastic scattering
 hinted at proton being a loosely bound compound object:
\begin{quote}
$\Longrightarrow$\qquad   {\em Constituent Quarks}
\end{quote}
 
\item
 Constancy of transverse momenta of produced hadrons, rare appearance
 of large-$k_\perp$ fluctuations, was signaling the weakness of
 interaction at small relative distances:
\begin{quote}
$\Longrightarrow$\qquad  {\em Asymptotic Freedom}
\end{quote}

\item 


 The total hadron interaction cross sections turned out to be
 practically {\em constant}\/ with energy. {\em If}\/ we were to
 employ the standard quantum field theory (QFT) picture of a particle
 exchange between interacting objects,
$$
   \sigma_{\mbox{\scriptsize tot}} \propto s^{J-1} \>\simeq\> \mbox{const},
$$
 {\em then} this called for a spin-one elementary field, $J=1$, to be
 present in the theory.

 {\em Uniformity in rapidity}\/ of the distribution of produced
 hadrons (Feynman plateau) pointed in the same direction, {\em if},
 once again, we were willing to link final particle production to
 accompanying QFT radiation.
\begin{quote}
$\Longrightarrow$\qquad  {\em Vector Gluons}.
\end{quote}
\end{itemize}
Nowadays the dossier of puzzles \&\ hints that the hadron
phenomenology has accumulated is very impressive. It includes a broad
spectrum of issues ranging from unexplained regularities in hadron
spectroscopy to soft ``forceless'' hadroproduction in hard processes.
%
%
To locate and formulate a puzzle, to digest a hint, -- these are the
road-signs to the hadron chromodynamics construction site.  We are
learning
to listen. And to hear.

\subsection{That nasty confinement}

The reason why one keeps talking, 30 years later, about {\em puzzles
and hints}, about {\em constructing}\/ QCD rather than {\em
applying}\/ it, lies in the conceptually new problem one faces when
dealing with a non-Abelian theory with unbroken symmetry (like
QCD). We have to understand how to master QFTs whose dynamics is
intrinsically unstable in the infrared domain: the objects belonging
to the physical spectrum of the theory (supposedly, colorless hadrons,
in the QCD context) have no direct one-to-one correspondence with the
fundamental fields the microscopic Lagrangian of the theory is made of
(colored quarks and gluons).

In these circumstances we don't even know how to formulate at the
level of the microscopic fields the fundamental properties of the
theory, such as conservation of probability (unitarity) and
analyticity (causality):
\begin{itemize}
\item
  What does {\em\bf Unitarity}\/ imply for confined objects?
\item
  How does {\em\bf Causality}\/ restrict quark and gluon Green functions
  and their interaction amplitudes?
\item
  What does the {\em\bf Mass}\/ of an INFO -- [well] Identified [but]
  Non-Flying Object -- mean?
\end{itemize}
The issue of quark masses is especially damaging since a mismatch
between quark and hadron thresholds
significantly affects predicting the yield of heavy-flavored hadrons
in hadron collisions. 

 Understanding the confinement of colour remains an open problem.
 Given the present state of ignorance, one has no better way but to
 circle along the {\em Guess-Calculate-Compare}\/ loop.  There are,
 however, {\em guesses}\/ and {\em guesses}.

\subsection{Circling the G-C-C loop}

 Perturbative QCD (pQCD) is believed to govern the realm of ``hard
 processes'' in which a large momentum transfer $Q^2$, either
 time-like $Q^2\gg 1$~GeV$^2$ (jets), or space-like $Q^2\ll
 -1$~GeV$^2$ (structure functions), is applied to hadrons.  pQCD
 controls the relevant cross sections and, to a lesser extent, the
 structure of final states produced in hard interactions.  Whatever
 the hardness of the process, it is hadrons, not quarks and gluons,
 that hit the detectors.  For this reason alone, the applicability of
 the pQCD approach, even to hard processes, is far from being obvious.
 One has to rely on plausible arguments (completeness, duality) and
 look for observables that are {\em less vulnerable}\/ towards our
 ignorance about confinement.

 Speaking of substituting {\em good guesses}\/ for {\em ignorance}\/
 the following ladder emerges.

\vspace{3mm}
\noindent
{\bf Total cross sections.}
 
 The safest bet of all is the idea of {\em Completeness}\/ applied to
 a handful of observables that enjoy the status of ``totally inclusive
 cross sections''. Completeness of colour states may be looked upon as
 a {\em good {\bf\em direct} guess}. The examples are
$$
 \sigma_{\mbox{\scriptsize tot}}(e^+e^-\to \mbox{hadrons}), \quad
 \Gamma(\tau \to \nu_\tau +  \mbox{hadrons}).
$$
Here one replaces the probability of production of {\em hadrons}\/ by
a colorless current $j$ (virtual photon, $Z^0$ or $W^\pm$) by that of
a $\qq$ pair,
\begin{equation}
\label{curr}
  W(j\to\mbox{hadrons}) \>=\> W(j\to\qq)\,\otimes\, \mbox{\bf 1}, 
\end{equation}
and argues that the {\em total}\/ probability of the conversion of
quarks into hadrons cannot be anything but {\bf 1}. This sounds fine
if the momentum transfer to the hadron system, $Q^2$, well exceeds the
hadron mass scale $\cO{1\GeV^2}$. The guess becomes less {\em
direct}\/ when the momentum transfer gets smaller and the final state
hadronic system starts to ``resonate''.  In particular, in the case of
the $\tau$ lepton decay width where $Q^2< m_\tau^2 \simeq
(1.8\,\GeV)^2$ a point-by-point correspondence between the left and
right hand sides of
\eqref{curr} is lost and some ``smearing'' over the invariant 
mass of the hadron system should be applied. There is a smart way to
do this, by referring to the {\em analyticity}\/ in $Q^2$ of the
correlator of the currents, $\lrang{j\,j}$, which follows from {\em
causality}. By treading this path one arrives at an amazingly tight
control over potentially disturbing non-perturbative effects, which
makes the $\tau$ decay a legitimate source of the $\as$ measurement at
pretty small scales (for details see~\cite{Beneke}).


\vspace{3mm}
\noindent
{\bf DIS structure functions.}

 These are not ``totally inclusive cross sections'', as far as hadrons
 are concerned, simply because there is a definite hadron in the {\em
 initial}\/ state.  We are not clever enough to
 deduce from first principles the parton distributions inside a target
 hadron (PDF, or structure functions). However, the rate of their $\ln
 Q^2$-dependence (scaling violation) is an example of a
 Collinear-and-Infrared-Safe (CIS) measure and stays under pQCD
 jurisdiction.  Here one applies a similar logic and appeals to
 analyticity of the virtual boson--proton scattering amplitude to
 translate the Bjorken-$x$ {\em moments}\/ of the inclusive
 Minkowskian DIS cross section (structure functions) into Euclidean
 space, the Operator Product Expansion (OPE) being the name of the
 game (a~{\em good {\bf\em indirect} guess}).

 Recall that in the Bjorken limit ($x\,$=~const, $|{Q^2}|\to\infty$,
 that is neglecting corrections in powers of $1/Q^2$) one can describe
 the pattern of the logarithmic deviations from the exact Bjorken
 scaling in terms of probabilistic QCD improved parton picture with
 cascading quarks and gluons replacing point-like partons of the
 original Bjorken-Feynman parton picture.

\vspace{3mm}
\noindent
{\bf Final state; Inclusive.}

 The next step down our squeaking ladder of ignorance -- and we arrive
 at inclusive characteristics of hadronic final states produced in
 hard processes. Oops! Here our guesses cannot be labeled other than
 {\bf\em wild}. There is no {\em a priory}\/ reason for distributions
 of final hadrons to bear much resemblance to those of underlying
 partons\footnote{modulo perturbatively
 controlled $Q^2$-dependence of the Feynman-$x$ moments of
 fragmentation functions, see below.}.  As we shall discuss below,
 both the energy and angular distributions of hadrons {\em do}\/
 follow partonic ones. This fact is well established
 phenomenologically (for a review see~\cite{KO}). It does not follow
 from ``first principles'', but rather tells us about confinement (as
 providing soft, local in the configuration space hadronization of
 partons) supporting the original wild guess known under the name of
 LPHD (Local Parton--Hadron Duality) hypothesis~\cite{LPHD}.

 It is important to mention that the probabilistic parton evolution
 picture (the source of inspiration for Monte Carlo event generators)
 is as {\em approximate}\/ as it is {\em limited}. Strictly speaking,
 it had been validated for DIS SFs~\cite{AP,GLD} (and single-particle
 inclusive distributions in $e^+e^-$ -- fragmentation
 functions~\cite{GLD}).  No less but no more. Aiming at more than that
 with MC tools is, strictly speaking, illegitimate. However this does
 not mean that a probabilistic treatment cannot be somewhat extended
 beyond its original limits.

 A famous example to the contrary is given by the so-called Angular
 Ordering (AO) story of early 1980s. Alfred Mueller and Victor Fadin
 found that quantum mechanical interferences affect soft gluon
 cascades in $e^+e^-$ jets and invalidate the classical (=
 probabilistic) DGLAP evolution picture~\cite{AP,GLD}. At the same
 time they showed that the interference effects could be taken full
 care of by simply restricting gluon multiplication into successively
 shrinking angular regions, $\Theta_{i+1}\ll \Theta_i$, with $i$ the
 parent parton and $i+1$ its softer offspring, $\omega_{i+1}\ll
 \omega_i$~\cite{AO}.  Surprizingly, the AO was later found to work
 beyond the leading strong ordering approximation (Double Logarithmic
 Approximation of strongly ordered energies {\em and}\/ angles,
 DLA). The most natural specification of the AO prescription,
 $\Theta_{i+1}\le \Theta_i$, was shown to properly embed the
 next-to-leading (single logarithmic, SL) corrections~\cite{LPHD}.
 This {\em Exact Angular Ordering}\/ rule (in place of the DL {\em
 Strong}\/ one) does the job and restores the probabilistic
 evolutionary picture for {\em energy}\/ spectra of (soft) particles
 in jets. As far as {\em angular}\/ distributions are concerned, it
 works, however, only {\em on average}. It cannot be applied to {\em
 angular correlations}. In particular, quantum-mechanical coherence
 plays a crucial r\^ole in predicting inter-jet particle flows in
 multi-jet events. This is the domain of the so-called string/drag
 phenomena, of collective radiation effects -- QCD radiophysics.

\vspace{3mm}
\noindent
{\bf Final state; Correlations.}

 Multi-particle correlations are obviously far more vulnerable. Even
 having learned and accepted the God\footnote{Uriel?}-given LPHD in
 single-particle distributions (inclusive particle flows), we
     feel at sea
 when correlations come onto stage. There may be some good news coming
 from pQCD approaches to KNO, intermittency phenomena and
 alike~\cite{DG}, but head-on perturbative attacks on correlations
 fail more often than~not.


\subsection{Substituting a good guess for ignorance}

 George Sterman and Steven Weinberg suggested to look for {\em
 Collinear-and-InfraRed-Safe}\/ (CIS) observables, those which can be
 calculated in terms of quarks and gluons without encountering either
 collinear (zero-mass quark, gluon) or soft (gluon) divergences.  
 They proclaimed such observables to be ``more equal'', free of large
 distance -- confinement -- effects and encouraged us to directly
 compare corresponding \PT\1 predictions with hadronic
 measurements~\cite{SW}.  This guess ranks higher than {\em
 hypothesis}: it is rather an {\em ideology}.

 The Sterman--Weinberg ideology gave rise to well elaborated
 procedures for counting jets (CIS jet finding algorithms) and for
 quantifying the internal structure of jets (CIS jet shape
 variables). They allow the study of the gross features of the final
 states while staying away from the physics of hadronization. Along
 these lines one visualizes asymptotic freedom, checks out gluon spin
 and colour, predicts and verifies scaling violation pattern in hard
 cross sections, etc. These and similar checks have constituted the
 basic QCD tests of the past two decades.

 This epoch is over. Now the HEP physics community aims at probing
 genuine confinement effects in hard processes to learn more about
 strong interactions.  The programme is ambitious and provocative.
 Friendly phenomenology keeps it afloat and feeds our hopes of
 extracting valuable information about physics of hadronization.
 The quest is not easy, we are bound to make mistakes and are trying
 to avoid errors.

\subsection{On mistakes vs.\ errors}

 \ford{0.52}{ mistake: \quad \begin{minipage}{0.9\textwidth} smth.\
 done wrongly, or \\ smth.\ that should not have been done.
 \end{minipage}\\[-5mm] 
 \begin{flushright} 
{\em Longman Dictionary of Contemporary English, 
 1987. 
} \end{flushright}}

 The original calculation of the electron loop which participates in
 the polarisation of QED vacuum and makes the coupling run with
 virtuality, $\alpha(k^2)$, produced a wrong sign (in modern words, a
 QCD-ish $\beta$-function).
 According to the Longman Dictionary this was an {\em error}, though
 not a {\em mistake}\/ since it was worth making!
 
 It took a while before some young colleagues\footnote{according to
 the legend, I.\ Tamm and A.\ Galanin~\cite{Dyatlov}.} pointed out to
 the ma\^{\i}tre that he erred. However the time span proved to be
 enough for Lev Landau to develop and enthusiastically discuss with
 Isaak Pomeranchuk and their pupils a beautiful physical picture of
 what is now known to us under the name of ``asymptotic freedom''.
 The seminal paper ``{\em On the quantum theory of fields}\/''
 followed~\cite{LAK}. It has the sign right; the {\em error}\/ had
 been corrected\footnote{This paper, however, contains a {\em
 mistake}\/ of a rather different nature. In a footnote we read:
 ``{\em The quadratically divergent photon mass should be put equal
 zero.}''  { [ So far so good. But then, ]} ``{\em The presence of a
 finite photon mass would violate the charge conservation law.}'' { [
 Nope. Though their footnote does fully apply to the QCD {\em
 gluon}}\/ mass. What an irony! ]}.

 In ``{\em Fundamental Problems}''~\cite{Landaulast} -- a homage to
 Wolfgang Pauli -- Landau discusses ``{\em `nullification' of the
 theory}\/'' which is ``{\em tacitly accepted even by theoretical
 physicists who profess to dispute it.}\/'' He remarks that ``{\em the
 validity of Pomeranchuk's proofs has been doubted}.'' He considers
 the criticism but asserts that ``{\em It therefore seems to me
 inappropriate to attempt an improvement in the rigor of Pomeranchuk's
 proofs, especially as the brevity of life does not allow us the
 luxury of spending time on problems which will lead to no new
 results.}''\footnote{This short but intensely wise
 paper turned out to be Landau's last.}

 In the late 1950s the problem was known as ``Moscow Zero'': vanishing
 of the physical interaction (renormalized coupling) in the limit of a
 point-like bare interaction, $\Lambda_{\mbox{\scriptsize
 UV}}\to\infty$.
 The depth of that crisis can be measured by the
 Dyson prophesy~\cite{Dyson} that the correct ``meson'' theory -- the
 theory of strong interactions -- ``{\em will not be found in the next
 hundred years}\/'' and/or by the
 Landau 
 conclusion~\cite{Landaulast}
 that ``{\em the Hamiltonian method for strong interactions is dead
 and must be buried, although of course with deserved honour.}''

 This was not an {\em error}. It was a {\em mistake}.  But one well
 grounded.
 It 
 was based on Pomeranchuk's extensive analysis of 
%
%
 all then-known renormalizable theories -- with scalar
 ($\lambda\phi^4$), Yukawa, four-fermion interactions~\cite{Pom}. In
 all these QFTs corresponding running couplings {\em increased}\/ with
 momentum transfer $|k^2|$, slowly but catastrophically.
 Let us not forget
 the same behaviour of QED~\cite{LAK} and an unrealistic but
 pedagogically valuable Lee model~\cite{KP}.
 No wonder, the situation looked desperate indeed.

 We may guess that Landau and Pomeranchuk apparently understood
 too well that search for a ``better'' (asymptotically free) theory
 was unlikely to bear fruit.
 The pattern of the fall-off ({\em screening}\/) of the interaction at
 {\em large distances}\/ (increase with momentum transfer) seemed too
 general to be passed by. Indeed, the vacuum polarisation loop
 corrections are analytic in $k^2$ (causality). Hence (by
 crossing-symmetry) the ``zero-charge'' sign of the $\beta$-function
 inevitably {\em follows}\/ from positivity of the cross-channel pair
 production cross section being proportional (by unitarity) to the
 imaginary part of the loop amplitude.

 Back in 1969 Yulik Khriplovich demonstrated that in a non-Abelian
 $SU(2)$ Yang--Mills gauge theory the coupling constant disrespects
 this argument~\cite{Yulik}. Vladimir Gribov
 explained how it dares to do so without violating ``first
 principles''~\cite{Anatomy}.
 
\noindent
\begin{minipage}{0.5\textwidth}
 Imagine a pair of static colour charges e.g., heavy quarks
 interacting via instantaneous Coulomb gluon exchange marked ``{0}''
 on the adjacent picture.

\parindent 3mm
 In the next order in $\as$ there appears the standard vacuum
 polarisation correction due to gluon decays into ``physical'' quanta,
 either a $\qq$ pair or two transverse gluons (``$\perp$'').  They
 both respect unitarity and give the same-sign contributions to the
 $\beta$-function as shown on the top part of the picture -- {\em
 screen}\/ the charge (as in QED and everywhere else). 

\parindent 3mm
 In QCD this is not the end of the story however. There is another
 type of radiative corrections due to the fact that our Coulomb
 carrier pro\-pagates in the ``external field'' of vacuum fluctuations
 of {\em transverse}\/ quanta. Coulomb gluons couple directly to
 transverse ones (whereas photons did not). The first non-vanishing
 contribution emerges in the second order 
\end{minipage} \qquad
\begin{minipage}{0.45\textwidth}
\begin{center}
  \epsfig{file=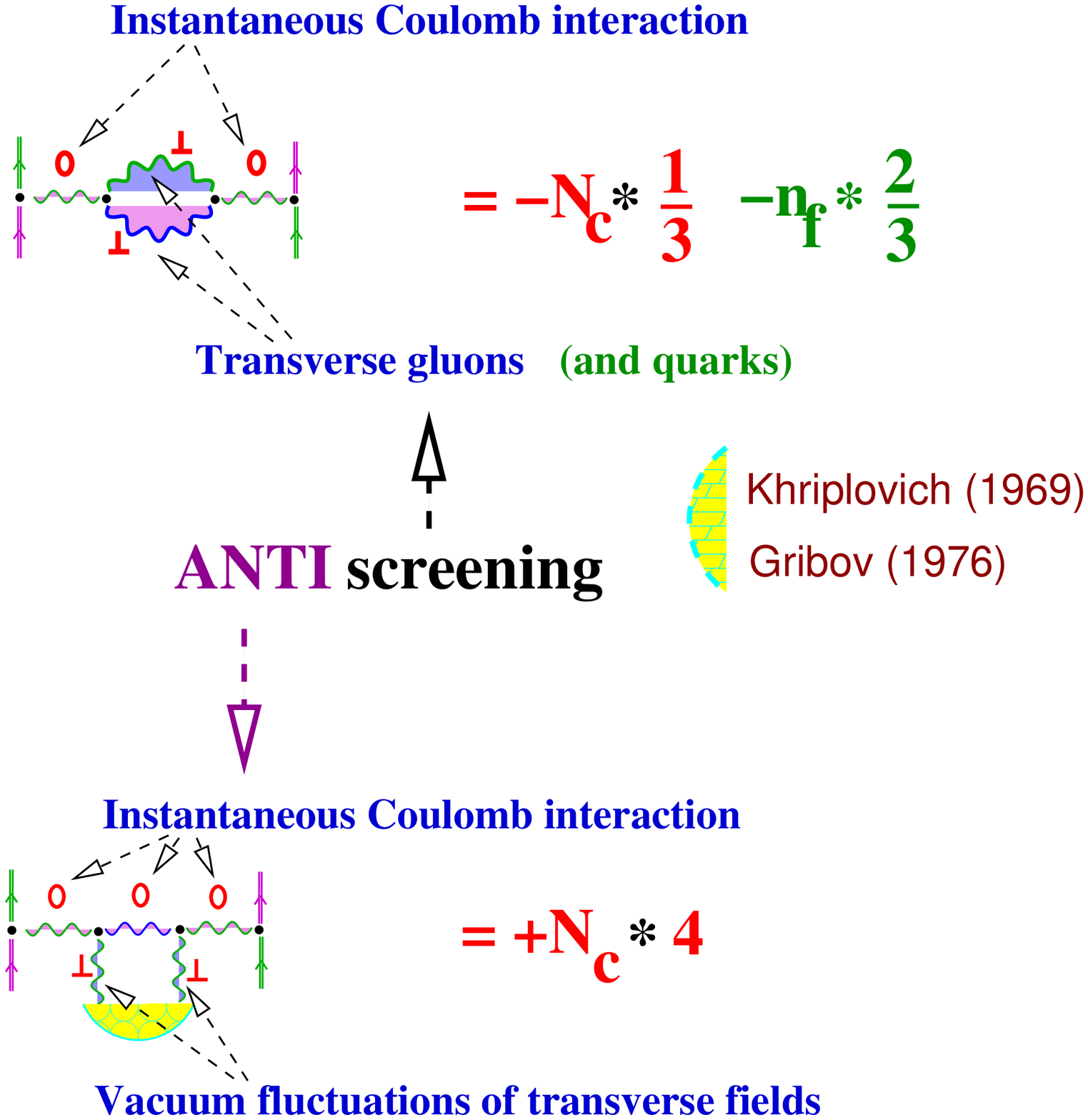,width=\textwidth}
\end{center}
\end{minipage}

\noindent
in the coupling $g_s$ (bottom part of the picture). It is
large and has the opposite sign corresponding to {\em
anti-screening}. The origin of the ``opposite sign'' is readily
understood~\cite{Anatomy}: it is the same phenomenon that pushes {\em
down}\/ the ground state energy of a quantum-mechanical system in the
second order in perturbation,
$$ 
    E_0'-E_0 = \sum_n \frac{\left|\lrang{0|\delta
V|n}\right|^2}{E_0-E_n} \> <\> 0.
$$

\subsection{Good guesses and bad guesses}

 Given our present awareness of
 the essential difference between mistakes and errors,
 an appeal 
 for 
 {\em bad guesses}\/ (BG) would not sound provocative.
 In the first place, one needs BGs to have {\em good guesses}\/ (GG)
 shining ever brighter. However to avoid slipping toward PR values we
 had better put forward a more serious argument: one learns by making
 bad guesses and confronting them with reality. QCD history is rich in
 BGs. Let us recall a couple of them.

 A celebrated example of a BG is given by the initial parton model
 picture of how quarks hadronize.
 Feynman's original idea was that each parton converts into a bunch of
 hadrons (with limited transverse momenta and a uniform distribution
 in $dx/x$) -- the Feynman {\em plateau}\/ of
 rapidity length $y_{\max}\simeq \ln E$, with $E$ the parton energy.
 This idea was realised in the very first (Field--Feynman)
 fragmentation model and was accommodated later by more advanced MC
 event generators like ISAJET, COJET. Such models today can be
 pronounced {\em dead}\/ and to {\em be buried, although of course with
 deserved honour}.
 They lost the race to the so-called Lund string model which was based
 on the smart decision to take into proper consideration the {\em
 colour topology}\/ of the underlying multi-parton
 system~\cite{Lund_string}.
  
 Bo Andersson, G\"osta Gustafson and Carsten Peterson\footnote{Theory
 Department of the Lund University, Sweden. Hence the name of the
 model.}
 chose to view a gluon radiated off a primary $\qq$ pair in
 $e^+e^-\to\qq$ as a system of a fake quark and antiquark,
 $q_f\bar{q}_f$.  Then with good accuracy (modulo $1/N_c^2
\sim 10\%$ colour-suppressed correction) each of the two ``pairs''
 $\bar{q}q_f$ and $\bar{q}_fq$ finds itself in a {\em colour
 singlet}\/ state. They suggested to treat each ``pair'' according to
 the Field--Feynman prescription (but in its proper cms!).
 This seemingly harmless modification had dramatic consequences.  For
 one thing, the multiplicity of additional hadrons originating from
 emission of a hard gluon turned out to be a function of the gluon
 {\em transverse momentum}\/ (with respect to the primary $\qq$)
 rather than its cms {\em energy}, $y_{\max}\simeq \ln k_\perp$. The
 crucial r\^ole of colour topology, both for multi-jet event
 multiplicities and for the pattern of particle flows between jets --
 the so-called string effect(s)~\cite{Lund_string}, was later
 confirmed by purely perturbative QCD considerations~\cite{drag}.

 Another prominent though less known example is provided by the story
 of the EEC (energy--energy correlation) measure in $e^+e^-$
 annihilation~\cite{EECdef}. It was the first CIS observable to
 have been experimentally studied at $e^+e^-$ accelerator PETRA in
 DESY (Hamburg). And with disastrous results.
 The ``ideology of infrared stability'' I was praising so above seemed
 to have failed. The discrepancy between the pQCD quark--gluon
 prediction and the measured hadron--hadron energy weighted inclusive
 correlation was found to be {\em substantial}. Worse than that, it
 turned out to be 
{\em  stubborn}\/ 
 as it refused to go away with increase of the annihilation energy
 $Q^2$, 
 defying ideology.
  
 Now we understand what has happened. The EEC in the back-to-back
 kinematics 
 turned out to be particularly strongly contaminated by
 non-perturbative effects~\cite{EEC}.

\section{PERTURBATIVE QCD AT WORK. WHY?}

\ford{0.6}{
 {\em He {\em [Dr.\ Dee]} deprecates any kind of traffic with
 unauthorised or unreliable spirits, and acknowledges again the only
 Source of wisdom. But since he has so long and faithfully followed
 learning, he does think it of importance that he should know
 more. The blessed angels, for instance, could impart to him things of
 at least as much consequence as when the prophet told Saul, the son
 of Kish, where to find a lost ass or two!}~\cite{char1}}

 In recent years pQCD has helped us to collect an impressive number of
 ``{\em lost asses}\/'' indeed. However one cannot help wondering why
 the pQCD treatment works so surprizingly well in some cases and fails
 miserably in others (often of a similar nature, residing on the same
 plank of our ladder of ignorance)?
 
 It seems the messages are being sent. To grasp them we have to
 separate, scrutinize and try to classify the ``good'' and ``bad''
 cases. But first we'd better agree on the vocabulary.

\subsection{Words, words, words \ldots}

Speaking of ``perturbative QCD'' can have two meanings.
\begin{itemize}
\item[\{1\}]
 In a narrow, strict sense of the word, {\em perturbative approach}\/
 implies representing an answer for a (calculable) quantity in terms
 of series in a (small) expansion parameter $\alpha_s(Q)$, with $Q$
 the proper hardness scale of the problem under consideration.

\item[\{2\}]
 In a broad sense, {\em perturbative}\/ means applying the language of
 quarks and gluons to a problem, be it of perturbative
 (short-distance, small-coupling) or even non-perturbative nature.
\end{itemize}

 The former definition \{1\} is doomed: the perturbative series so
 constructed are known to diverge. In QCD these are asymptotic series
 of a kind that cannot be ``resummed'' into an analytic function in a
 unique way. For a given calculable (collinear-\&\ {}-infrared-safe;
 CIS) observable~\cite{SW} the nature of this nasty divergence can be
 studied and quantified as an intrinsic uncertainty of pQCD series, in
 terms of so-called {\em infrared renormalons}\/~\cite{Beneke}.  Such
 uncertainties are non-analytic in the coupling constant and signal
 the presence of non-perturbative (large-distance) effects.  For a CIS
 observable, non-perturbative physics enters at the level of
 power-suppressed corrections $\exp\{-c/\alpha_s(Q)\}\propto Q^{-p}$,
 with $p$ an observable-dependent positive integer\footnote{usually,
 though not necessarily~\cite{EEC}} number.

 Meanwhile the broader definition \{2\} of being ``perturbative'' is
 bound to be right. At least as long as we aim at eventually deriving
 the physics of hadrons from the quark-gluon QCD Lagrangian.

 To distinguish between the two meanings, in what follows we will
 supply the word {\em perturbative}\/ with a superscript \1 or \2.
 Thus, when discussing the strong interaction domain in terms of
 quarks and gluons in what follows we will be actually speaking about
 perturbatively\1 probing non-perturbative\1 perturbative\2 effects.

\subsection{QCD coupling}
\begin{flushright}
\begin{minipage}{0.5\textwidth}
{\em Loke unto thy charge truely: \\
    Thow art yet dead. Thow shallt be revyved.}\/~\cite{chm}
\end{minipage}
\end{flushright}

\begin{center}
\epsfig{file=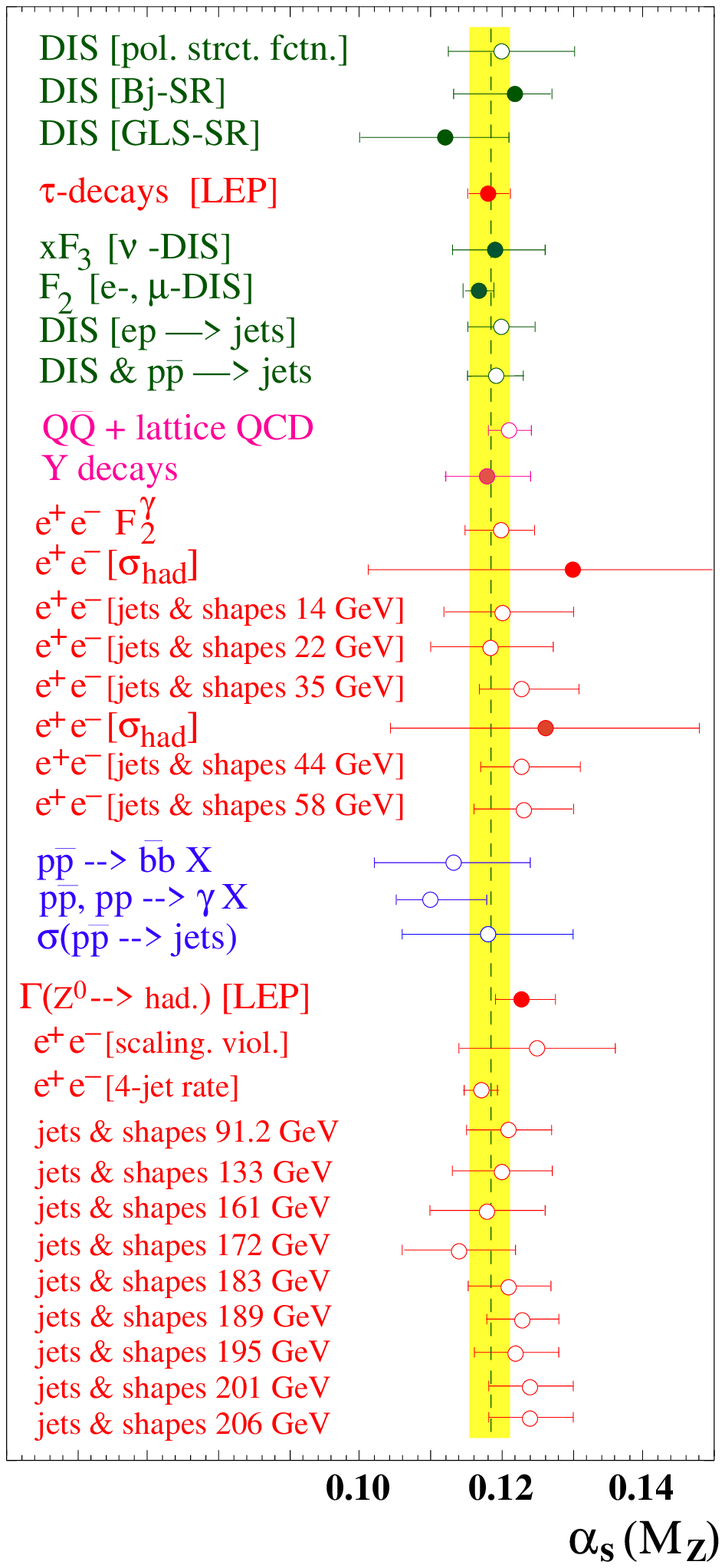,width=0.4\textwidth,height=0.7\textwidth}
\hfill
\begin{minipage}[b]{0.55\textwidth}
 The pictures of the properly measured and properly running
 perturbative\1 QCD coupling~\cite{ZB} are soothing.

\vfill

\noindent
\epsfig{file=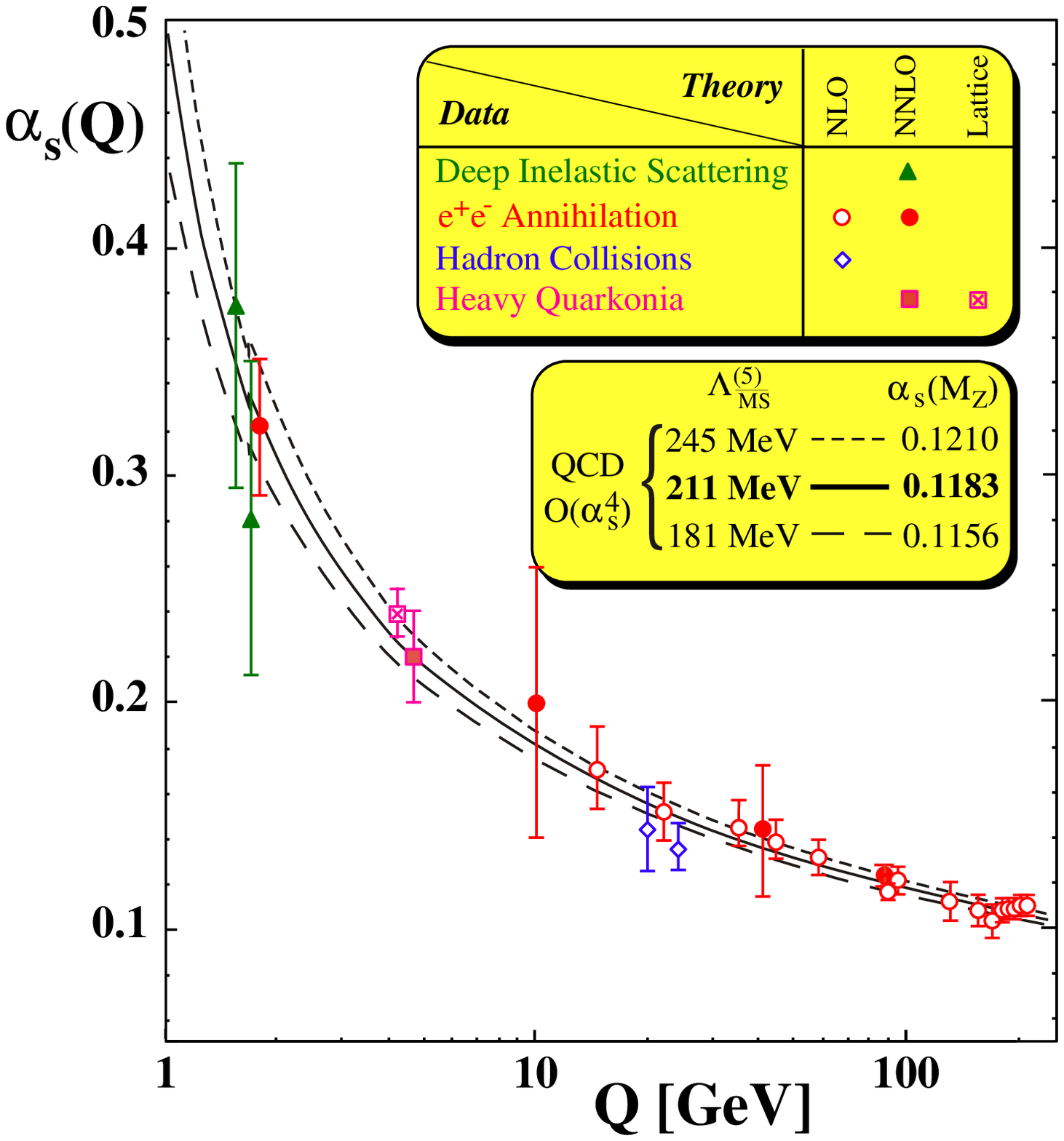,width=\textwidth,height=1.15\textwidth}
\end{minipage}
\end{center}

 Experimenters (as well as theorists) shy away from looking below
 1~\GeV. And for a good reason too: how to discuss the strength of
 interaction between colored objects -- quarks and gluons -- that
 supposedly ``don't exist'' at ``large'' distances
 corresponding to $Q<1\,\GeV$?  We may eventually have to.

 Recall what the Renormalization Group teaches us about the change of
 renormalized coupling $\as(\mu_R^2)$ with the renormalization scale
 $\mu_R$.  This teaching, however, is of limited value.  The momentum
 variation of $\as$ is determined by the $\beta$-function,
$$ 
 \frac{d}{d\ln\mu_R} \left(\frac{\as(\mu_R^2)}{2\pi}\right)^{-1} =
 \beta(\as(\mu_R^2))\>, \qquad
 \beta(\al) \>=\> \beta_0 + \beta_1\frac{\al}{2\pi}  + \ldots \,.
$$
 Beyond two loops the coefficients $\beta_{n}$ with $n\ge2$ turn out
 to be {\em scheme}- (and in some schemes even {\em gauge})-dependent;
 in other words, arbitrary.  Therefore, the large-momentum behaviour
 of the running coupling $\as(Q^2)$ cannot be uniquely fixed beyond
 two loops. The reason for that is pretty simple.  Universality is
 inherited from the basic property of {\em ultraviolet
 renormalizability}\/ of the theory, and it is only the first two
 loops that are truly dominated by the UV region, by small-distance
 physics.

 Indeed, the one-loop radiative corrections contain the standard
 logarithmically divergent integral
$$
\int \frac{d^4q}{q^4}\propto \ln\Lambda_{\UV} 
\>\>  =\>\> \infty
\quad {\Longrightarrow} \quad \cb_0\>.
$$
Hiding infinity under the carpet produces $\cb_0$, the first
coefficient in the \PT\ $\cb$-function expansion. In the next step we
supply our loop with an additional internal gluon. Now we have two
independent loop-momenta to integrate over, $q_1$ and
$q_2$. Integration regions $q_{1(2)}\ll q_{2(1)}\ll \Lambda_\UV$ could
have produced $(\ln\Lambda_\UV)^2$ contributions. These get suppressed
by renormalizing the {\em internal}\/ propagators and vertices at the
one-loop level, the result being a single-logarithmic integral
determined by the region $q_1\sim q_2\ll \Lambda_\UV$,
\begin{equation}
  \label{eq:oneloop}
\int \frac{d^4q}{q^4}\as(\mu_R^2)\>\>\propto\>\> 
\as(\mu_R^2) \ln\Lambda_{\UV}\>\>=\>\> \infty 
\quad {\Longrightarrow} \quad \cb_1\,.  
\end{equation}
This is how the usual story goes, order by order in perturbation theory. 
We can do better, however, by taking into consideration that the coupling in
\eqref{eq:oneloop} runs with the internal momentum.  This means
reorganising the \PT\ series so as to incorporate into the two-loop
diagram the higher order effects which result in substituting the
running $\as(q^2)$ for the constant $\as(\mu_R^2)$. By doing so we
obtain a contribution which is still \UV-divergent, though modified by
the logarithmic decrease of the coupling at large momenta:
$$
 \int \frac{d^4q}{q^4}\as(q^2)\propto 
 \ln\ln\Lambda_{\UV}\>\> =\>\> \infty 
 \quad {\Longrightarrow} \quad \cb_1\,.
$$
Renormalizing it out gives rise to $\cb_1$. Starting from the third
loop (twointernal gluons) the situation however changes drastically:
the \UV-region is no longer dominant, and we get
$$
 \int \frac{d^4q}{q^4}\left(\as(q^2)\right)^2\>\>=\>\> \mbox{finite}
\quad{\Longrightarrow} \quad \mbox{
$\cb_{n\ge2}$ depend on the {\bf infrared} physics!}  
$$
 Thus starting from the $\as^3$ (next-to-next-leading) level, a purely
 perturbative\1 treatment may become intrinsically ambiguous because
 of an interconnection between small and large distances. There is no
 way of unambiguously defining the QCD coupling $\as$ (beyond two
 loops) without solving the Theory in the infrared, that is without
 understanding the physics of colour confinement.

\subsection{Where is confinement?}

 The quark--gluon picture works rather well across the board. 
 Moreover, in many cases it seems to work {\em too well}. This is
 another worry: too good to be true ain't good enough.

\vspace{3mm}
\noindent
{\bf Too early?}

 The way the differential large angle $2\to2$ particle scattering
 cross sections should scale with energy (momentum transfer) was
 envisaged by the so-called ``quark counting rules''~\cite{MMT},

$$
   \frac{d\sigma}{dt} = \frac{f(\Theta)}{s^{{K}-2}} ; \qquad
   \frac{t}{s} = \mbox{const},
$$
 with $K$ the number of {\em elementary fields}\/ (quarks, photons,
 leptons, etc.) among~/~inside the initial and final particles. 

 For example, in the case of the deuteron break-up by a photon,
 $\gamma + D\to p+n$, we have $K=1+6+6=13$ (a photon and 6 quarks
 inside the initial deuteron and another 6 in the final proton and
 neutron). So, the differential cross section is expected to fall with
 $s$, {\em asymptotically}, as $s^{-11}=E_{\mbox{\scriptsize
 c.m.}}^{-22}$.  The key word {\em asymptotically}\/ always provided
 an excuse for unnerved HEP theorists in their encounters with angered
 experimenters.  The JLAB plot in Fig.~1 which I borrowed from Paul
 Hoyer's talk~\cite{Hoyer} seems to be telling us that this standard
 excuse is unnecessary here. However, it is again unnerving but for
 precisely opposite reason, if you take my meaning. Indeed, it is {\em
 very difficult}\/ to digest how the naive asymptotic regime manage to
 settle {\em that}\/ early! The lab.\ energy $1\,\GeV$ of the incident
 photon, where the scaling behaviour starts, is just {\em too}\/ low.

\noindent
\begin{minipage}{0.4\textwidth}
 The ``counting rules'' invite us to view a fast deuteron as a system
 of six comoving valence quarks. One of them is punched by the
 photon. The other five we have to properly push ourselves so as to
 make them fit into two outgoing nucleons. This is done by exchanging
 five gluons between the quarks in the {\em scattering amplitude}\/ so
 that the {\em cross section}\/ acquires the factor $\as^{10}$.  The
 picture makes sense as long as 1)~the deuteron is indeed {\em fast}\/
 and 2)~typical momentum transfers $q^2$ between quarks are large
 enough to allow us to use the concept of gluon exchange and of the
 QCD\1 coupling $\as(q^2)$ for that
\end{minipage}\hfill
\begin{minipage}{0.55\textwidth}
\epsfig{file=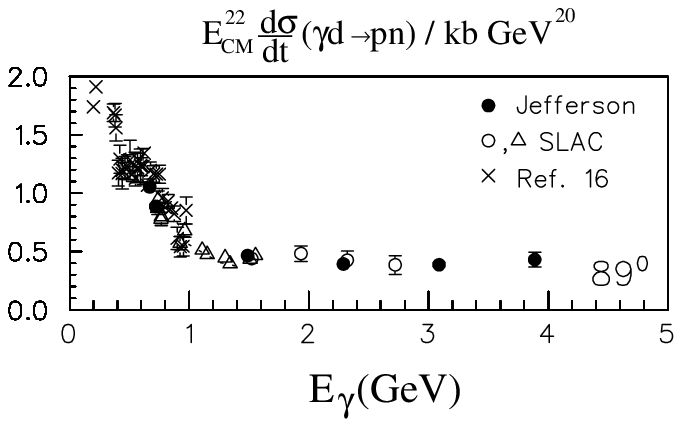,width=\textwidth,height=0.73\textwidth}

Fig.~1: Large angle $\gamma$-disintegration of a deuteron~\cite{JLABdata}.
\end{minipage}

\noindent
matters. None of these conditions holds for $E_\gamma\simeq
1\,\GeV$. 

Nonetheless we would have had every right to feel happy about Fig.~1
provided we could convincingly answer but one question: why is such
precocious scaling not seen for simpler systems and in particular for
the simplest of them all -- the electromagnetic form factor of a pion?

\vspace{3mm}
\noindent
{\bf Too smooth?}

\noindent
\begin{minipage}{0.5\textwidth}
 HERA measurements of the DIS proton structure function $F_2(x,Q^2)$
 in a wide range of photon virtualities,
$$ 
    0.1\,\GeV^2< Q^2 < 35\,\GeV^2,
$$ 
are compiled in Fig.~2.  The data are plotted as a function of the
simple variable 
$$
\xi=\log\frac{0.04}{x} \log\left(1+\frac{Q^2}{0.5\,\GeV^2}\right)
$$
 proposed by Dieter Haidt \cite{Haidt}.

\parindent 3mm
 Being surprisingly smooth, they show no sign of a ``phase
 transition'' when going from large virtualities (perturbative\1
 regime) downto very small scales where non-perturbative\1 physics
 should dominate.
\end{minipage}
\begin{minipage}{0.5\textwidth}
\begin{center}
  \includegraphics[width=\textwidth]{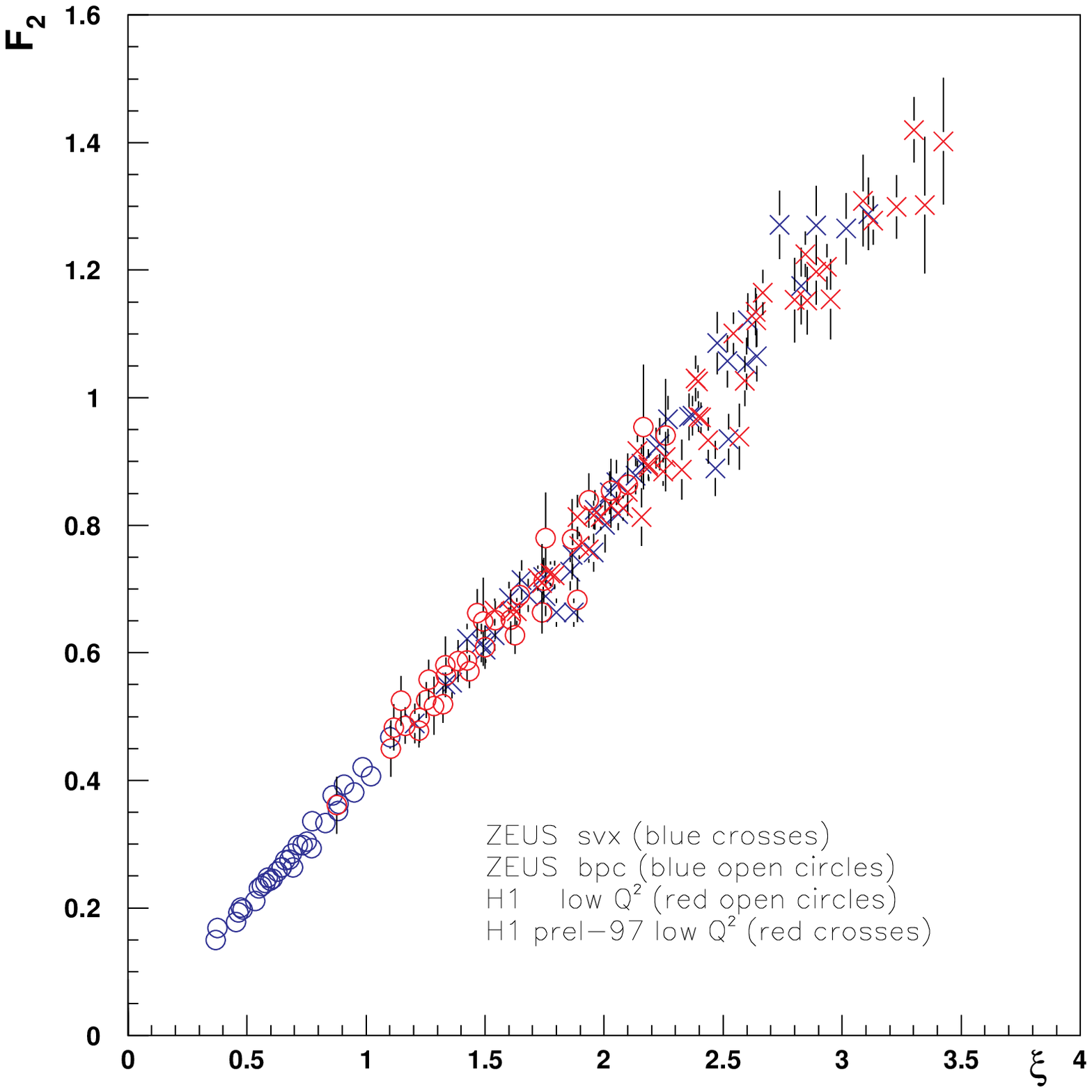} 

Fig.~2: $F_2$ for $x\leq 10^{-3}$,  $Q^2 \ge 0.1\,\GeV^2$~\cite{Haidt}.  
\end{center}
\end{minipage}


\vspace{3mm}
\noindent
{\bf Too soft?}

  As we will discuss below in great detail, the perturbatively\1
  predicted inclusive energy spectra of relatively soft, $x_p\ll 1$,
  {\em partons}\/ (mostly gluons) were found at LEP, HERA, Tevatron
  and elsewhere to be {\em mathematically similar}\/ to those of
  charged hadrons (mostly pions), thus confirming the LPHD hypothesis
  of soft confinement.  The \PT\ distribution in $\ln 1/x_p$ has a
  characteristic shape which follows from coherent gluon cascades
  (abovementioned AO). The predicted position of the hump for gluons
  coincides with the maximum of the pion spectrum and lies, typically,
  below $p=1~\GeV$!

  The same story with angular distributions of interjet soft particle
  flows in multi-jet ensembles (numerous string/drag effects). The
  worry is, that these in-between jets particles are in reality but
  $100--300\,\MeV$ pions which for some reason beyond our apprehension
  still choose to obediently follow the pattern of underlying colour
  fields. The message is strange but clear: whatever the ultimate
  solution of the confinement problem may be, it had better be gentle
  in transforming the quark-gluon Poynting-vector into the
  Poynting-vector of the final state hadrons.

\vspace{3mm}
\noindent
{\bf Is proton really bound?}

 HERA taught us that proton is fragile. It suffices to kick it with
 1~\GeV\ momentum transfer, or even less, to blow it to pieces.  It
 seems that what keeps a proton together is not any strong forces
 between the quarks but merely quantum mechanics: the proton just
 happened to be the ground state with a given well conserved quantum
 number (baryon charge). It is interesting to see how easy it is to
 break a proton. To achieve that it is not even necessary to kick it
 hard. A soft scratch (or rather two) is enough to do the job.

\noindent
\begin{minipage}{0.5\textwidth}
\epsfig{file=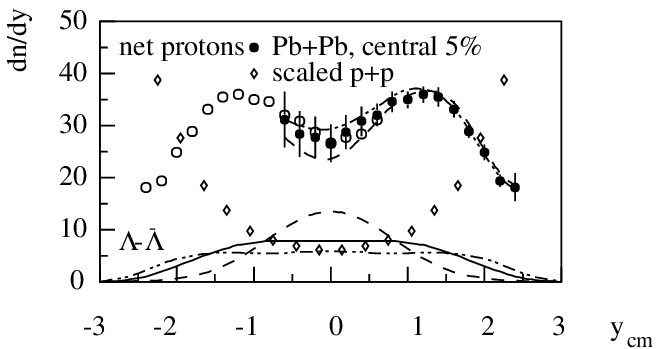,width=1.05\textwidth,height=0.6\textwidth}
\vspace{-3mm}
{Fig.~3: Proton ``stopping'' as seen by NA-49 (1999)}
\end{minipage}
\hfill
\begin{minipage}{0.48\textwidth}
 There is no sign of advocated fragility in a normal (minimum bias,
 soft) high energy proton-proton scattering.  The famous leading
 particle effect shows that a projectile proton stays intact in the
 final state and carries away a major fraction of the incident
 momentum (diamonds for ``scaled $p+p$'' in Fig.~3).
 This should not surprise us. In a typical $pp$ interaction it is only
 one of the valence quarks of the proton that scatters. Internal
 coherence of the spectator quark pair remains undisturbed. In these
 circumstances the proton splits into a triplet quark and a spectator
 diquark which is in a colour 
\end{minipage}

\vspace{3mm}

\noindent
\begin{minipage}{0.45\textwidth}
 anti-triplet state.  At the hadronization stage, the former picks up
 an antiquark and turns into a meson carrying, roughly,
 $z\simeq\frac13$ of the initial proton momentum, while the diquark
 (colour equivalent of a $\bar{q}$) picks up a quark forming a leading
 baryon with $z\simeq\frac23$. It may be, for example, a
 $\Lambda$--baryon as shown in Fig.4a. More often it will be a proton,
 neutron or $\Delta$. What is important, however, is that the {\em
 baryonic}\/ quantum number moves forward -- stays close in rapidity
 to the projectile proton.
\end{minipage} \quad
\begin{minipage}{0.5\textwidth}
\epsfig{file=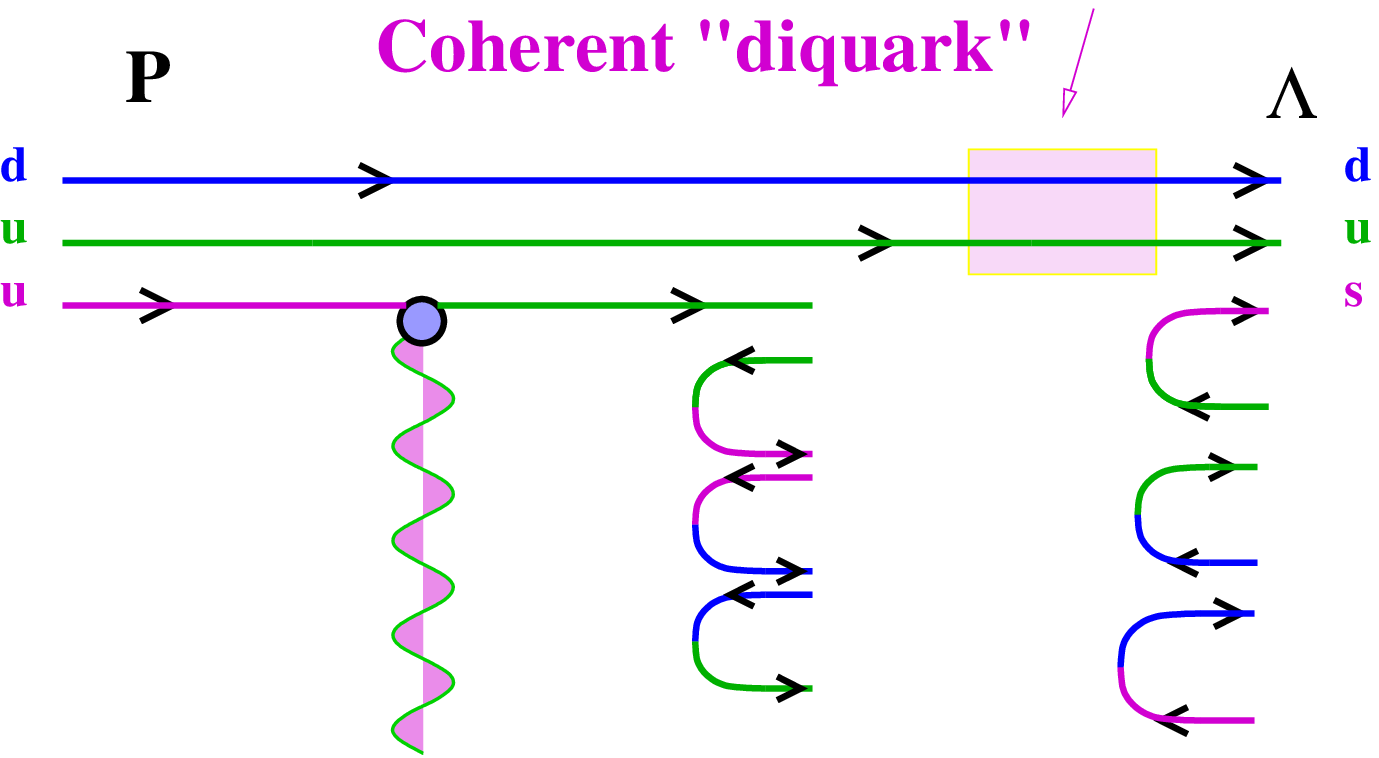,width=\textwidth}
Fig.~4a:  \hfill Gluon exchange produces a leading baryon.
\end{minipage}

\noindent
\begin{minipage}{0.45\textwidth}
\parindent 3mm
 It suffices, however, to organize a {\em double}\/ scattering within
 a life-time of the intrinsic proton fluctuation in order to destroy
 the proton coherence completely (including that of the diquark which
 remains intact after the first scratch).  Now the three
 quark-splinters of the proton separate as independent triplet charges
 and normally convert in the final state into three leading mesons
 carrying $z\simeq\frac13$ each as Fig.~4b suggests, with the baryon
 quantum number sinking into the sea.
%
%
\end{minipage}
\hfill
\begin{minipage}{0.5\textwidth}
\epsfig{file=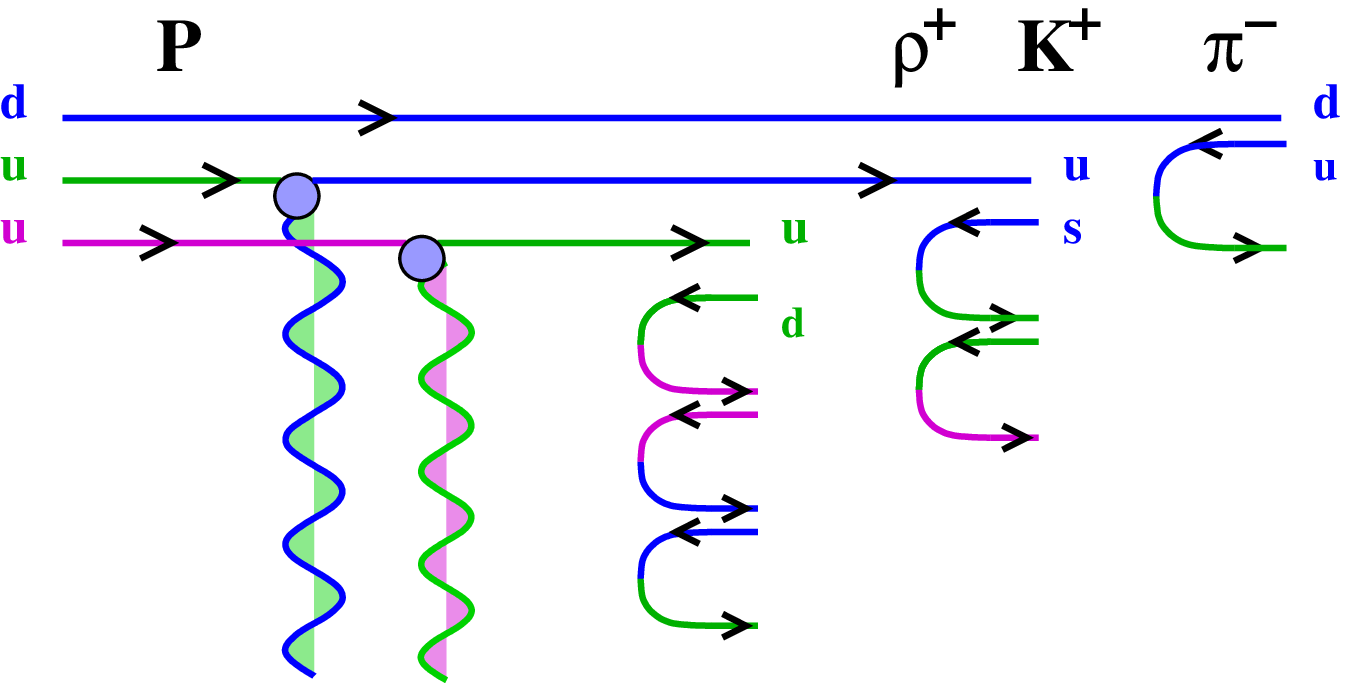,width=\textwidth}
Fig.~4b:\hfill Double exchange ``breaks up'' the proton.
\end{minipage}

\noindent

 This is what seems to be going on in the lead--lead scattering, see
 Fig.~3. Disappearance of leading protons is known as ``stopping'' in
 the literature.  This I believe is an inadequate name: there is no
 way to {\em stop}\/ an energetic particle, especially in soft
 interaction(s).  Relativistic quantum field theory is more tolerant
 to {\em changing particle identity}\/ than to allowing a large
 transfer of energy-momentum (recall relativistic Compton where the
 {\em backward}\/ scattering dominates: an electron turns into a
 forward photon, and vice versa).

 If this heretic explanation of the ``stopping'' as proton instability
 is correct, the same phenomenon should be seen in the proton
 hemisphere of proton-nucleon collisions and even in $pp$.  As we
 know, in $pp$ there are leading protons. However, this is true {\em
 on average}.  Even in $pp$ collisions one can enforce multiple
 scattering (and thus full proton breakup) by selecting rear events,
 e.g.\ with larger than average final state multiplicity.

 In all these cases ($pp$, $pA$, $AB$) ``proton decay'' should be
 accompanied by an enhanced strangeness production. 

\subsection{Perturbative\1 quark confinement?}

 \ford{0.58} {{\em A spirit
 afterwards told him {\em [John Dee]} that ignorance was the nakedness
 wherewith he was first tormented, and ``the first plague that fell
 unto man was the want of science.''}~\cite{char1}} 

 Soft hadronization, likely absence of strong inter-parton forces,
 fragile proton -- can it be reconciled with confinement in the first
 place?
To the best of my knowledge, the Super-Critical Light-Quark
Confinement theory (GSCC) suggested by V.N.~Gribov in early
90s~\cite{GSCC} is the only scenario to offer a natural explanation to
the puzzling phenomenology of multi-hadron production discussed above.

As a result of the search for a possible solution of the confinement
puzzle Gribov formulated for himself the key ingredients of the
problem and, correspondingly, the lines to approach it:
\begin{itemize}
\item
    The question of interest is not of ``a'' confinement, but that of
    ``the'' confinement in the real world, namely, in the world with
    two very light quarks ($u$ and $d$) whose Compton wave lengths are
    much larger than the characteristic confinement scale ($m_q\sim
    5-10\,\MeV \ll 1\,\GeV$).
\item
    No mechanism for binding massless {\em bosons}\/ (gluons) seems to
    exist in QFT, while the Pauli exclusion principle may provide
    means for binding together massless {\em fermions}\/ (light
    quarks).
\item
    The problem of ultraviolet regularization may be more than a
    technical trick in a QFT with apparently infrared-unstable
    dynamics: the ultraviolet and infrared regimes of the theory may
    be closely linked. Example: the pion field as a Goldsone boson
    emerging due to spontaneous chiral symmetry breaking (short
    distances) and as a quark bound state (large distances).
\item
    The Feynman diagram technique has to be reconsidered in QCD if one
    goes beyond trivial perturbative correction effects.  Feynman's
    famous $i\epsilon$ prescription was designed for (and is
    applicable only to) the theories with stable perturbative vacua.
    To understand and describe a physical process in a confining
    theory, it is necessary to take into consideration the response of
    the vacuum, which leads to essential modifications of the quark
    and gluon Green functions\footnote{The proper technology lies in a
    generalisation of the Keldysh diagram technique designed to
    describe kinetics out of equilibrium.}.
\end{itemize}

\noindent
There was a deep reason for this turn, which Gribov formulated in the
following words:
\begin{quote} 
  ``I found I don't know how to bind massless {\em bosons}''
\end{quote}
(read: how to dynamically construct {\em glueballs}\/).  

As for fermions, there is a corresponding mechanism provided by the
Fermi-Dirac statistics and the concept of the ``Dirac sea''.  Spin-$\frac12$
particles, even massless which are difficult to localize, can be held
together simply by the fact that, if pulled apart, they would
correspond to the free-fermion states that are {\em occupied}\/ as
belonging to the Dirac sea.

Thus, light quarks are crucial for GSCC. 
%
%
It is clear without going into much mathematics that the presence of
light quarks is sufficient for preventing the colour forces from
growing real big: dragging away a heavy quark we soon find ourselves
holding a blanched $D$-meson instead. The light quark vacuum is eager
to screen any separating colour charges.

The question becomes quantitative: how strong is strong? How much of a
tension does one need to break the vacuum and organize such a
screening?  

In a pure perturbative (non-interacting) picture, the empty fermion
states have {\em positive energies}, while the {\em negative-energy}\/
states are all filled.
With account of interaction the situation may change, {\em provided}\/
two {\em positive-energy}\/ fermions (quarks) were tempted to form a
bound state with a {\em negative}\/ total energy. 
In such a case, the true vacuum of the theory would contain {\em
positive kinetic energy}\/ quarks hidden inside the {\em negative
energy}\/ pairs, thus preventing positive-energy quarks from flying
free.

A similar physical phenomenon is known in QED under the name of
super-critical binding in ultra-heavy nuclei.
Dirac energy levels of an electron in an external static field created
by the large point-like electric charge $Z>137$ become {\em
complex}. This means instability.  Classically, the electron ``falls
onto the centre''. Quantum-mechanically, it also falls, but into the
Dirac sea.

In QFT the instability develops when the energy $\epsilon$ of an empty
atomic electron level falls, with increase of $Z$, below $-m_ec^2$.
An $e^+e^-$ pair pops up from the vacuum, with the vacuum electron
occupying the level: the super-critically charged ion decays into an
``atom'' (the ion with the smaller charge, $Z-1$) and a real positron
$$
 A_Z \>\Longrightarrow A_{Z-1} + e^+\,, \qquad \mbox{for}\>
 Z>Z_{\mbox{\scriptsize crit.}}
$$ 
Thus, the ion becomes unstable and gets rid of an excessive electric
charge by emitting a positron~\cite{PM45} 
%
 In the QCD context, the increase of the running quark-gluon coupling
 at large distances replaces the large $Z$ of the QED problem.

 Gribov generalised the problem of super-critical binding in the field
 of an infinitely heavy source to the case of two massless fermions
 interacting via Coulomb-like exchange.  He found that in this case
 the super-critical phenomenon develops much earlier. Namely, a {\em
 pair of light fermions}\/ interacting in a Coulomb-like manner
 develops super-critical behaviour if the coupling hits a definite
 critical value
\begin{equation}
\label{eq:acr}
 \frac{\alpha}{\pi} > \frac{\acr}{\pi} = 1-\sqrt{\frac{2}{3}}\,.
\end{equation}
 In QCD one has to account for the colour Casimir operator. Then the
 value of the coupling above which restructuring of the \PT\ vacuum
 leads to chiral symmetry breaking and, likely, to confinement
 (\cite{GSCC} and references therein), translates into
\begin{equation}
\label{eq:acrQCD}
  \frac{\acr}{\pi} = C_F^{-1}\left[\, 1-\sqrt{\frac{2}{3}}\,\right]
  \simeq 0.137\,.
\end{equation}
 This number, apart from being easy to memorize, has another important
 quality: it is numerically small. Gribov's ideas, being understood
 and pursued, offer an intriguing possibly to address all the
 diversity and complexity of the hadron world from within the field
 theory with a reasonably small effective interaction strength (read:
 not only {\em perturbatively}\2 but {\em perturbatively}\1).

\section{MULTIPLE HADROPRODUCTION: ASCENDING THE LADDER}

 We have already carefully measured our steps 
 down the ladder of ignorance.  Now let us ascend another one, looking
 for indirect and then direct evidences in favour of
 quark-gluon dynamics in multiple hadroproduction.
  
 High energy $e^+e^-$ annihilation, DIS, production in hadron-hadron
 collisions of massive lepton pairs, heavy quarks and their bound
 states, of large transverse momentum jets and photons are classical
 examples of hard processes.  Copious production of hadrons is typical
 for all of them.  On the other hand, at the microscopic level,
 multiple quark-gluon ``production'' is to be expected as a result of
 QCD bremsstrahlung -- gluon radiation accompanying abrupt
 creation/scattering of colour partons.

\subsection{Scaling violation pattern (indirect evidence)}

 Indirect evidence that gluons are there, and that they behave, can be
 obtained from the study of the scaling violation pattern.  QCD quarks
 and gluons are not point-like particles, as the orthodox parton model
 once assumed. Each of them is surrounded by a proper field coat -- a
 coherent virtual cloud consisting of gluons and ``sea'' $q\bar{q}$
 pairs.  A hard probe applied to such a dressed parton breaks
 coherence of the cloud. Constituents of these field fluctuations are
 then released as particles accompanying the hard interaction.

 The harder the hit, the larger an intensity of bremsstrahlung and,
 therefore, the fraction of the energy-momentum of the dressed parton
 that the bremsstrahlung quanta typically carry away.
 Thus we should expect, in particular, that the probability that a hit
 ``bare'' core quark carries a large fraction $x\sim1$ of the energy
 of its dressed parent will decrease with increase of $Q^2$.  And so
 it does.
 The logarithmic scaling violation pattern in DIS structure functions
 is well established and meticulously follows the QCD prediction based
 on the parton evolution picture.

 In DIS we look for a ``bare'' quark inside a target dressed one. In
 $e^+e^-$ hadron annihilation at large energy $s=Q^2$ the chain of
 events is reversed.  

 Here we produce instead a bare quark with energy $Q/2$, which then
 ``dresses up''.  In the process of restoring its proper field-coat
 our parton produces (a controllable amount of) bremsstrahlung
 radiation which leads to formation of a hadron jet.  Having done so,
 in the end of the day it becomes a constituent of one of the hadrons
 that hit the detector.  Typically, this is the leading
 hadron. However, the fraction $x$ of the initial energy $Q/2$ that is
 left to the leader depends on the amount of accompanying radiation
 and, therefore, on $Q^2$ (the larger, the smaller). 

 In fact, the same rule (and the same formula) applies to the scaling
 violation pattern in $e^+e^-$ fragmentation functions (time-like
 parton evolution) as to that in the DIS parton distributions
 (space-like evolution).

\noindent
\begin{minipage}{0.5\textwidth}
\epsfig{file=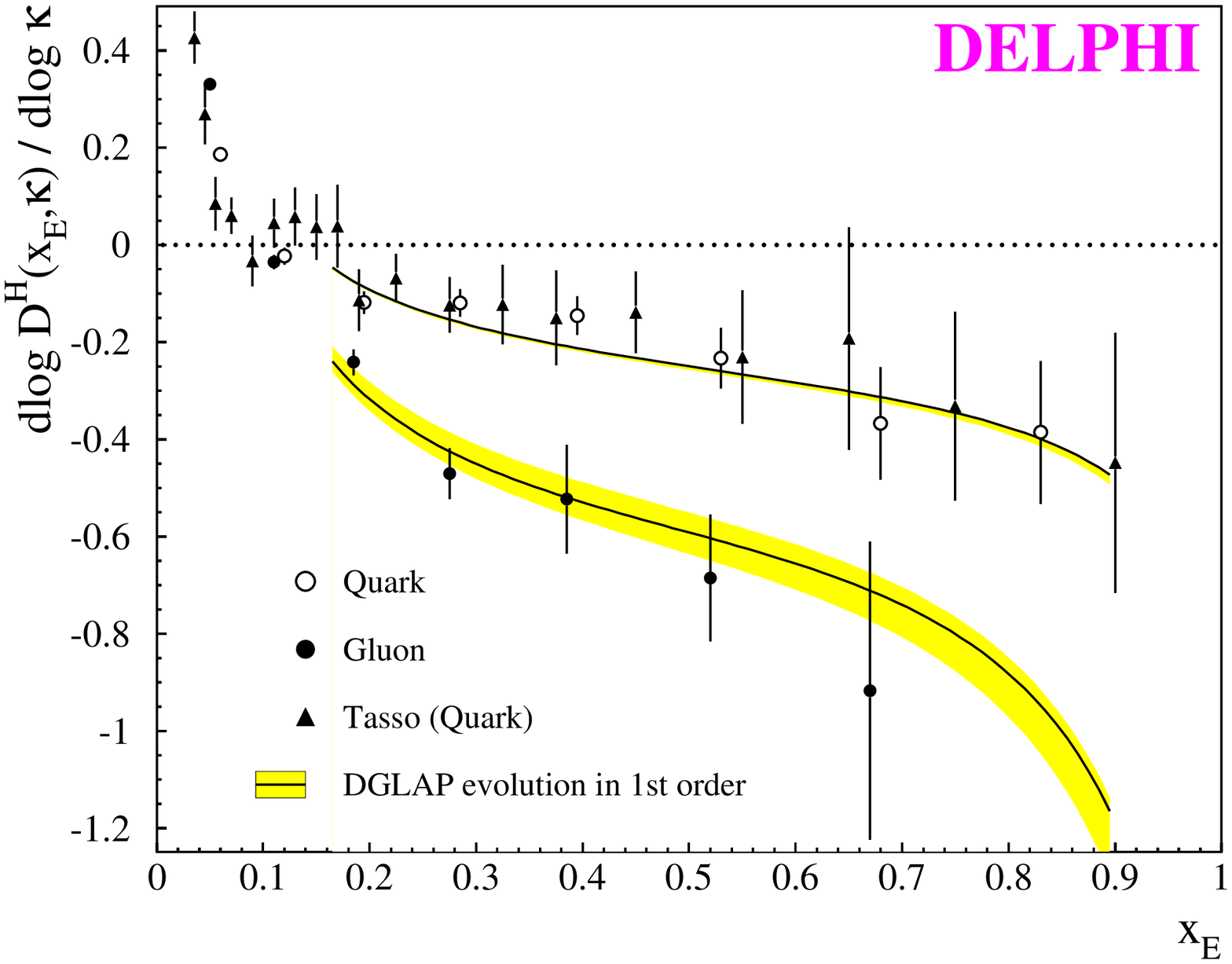,width=\textwidth}

Fig.~5: Comparison of scaling violation 
 in inclusive hadron distributions from gluon and quark jets
 {\protect\cite{Abreu}.}
\end{minipage}
\hfill
\begin{minipage}{0.45\textwidth}
\parindent 3mm
 The $e^+e^-$ annihilation experiments have become so sophisticated as
 to provide us with a near-to-perfect separation between quark- and
 gluon-initiated jets (the latter being extracted from
 heavy-quark-tagged three-jet events).

\parindent 3mm
 In Fig.~5 a comparison is shown of the scaling violation rates in the
 hadron spectra from gluon and quark jets, as a function of the
 hardness scale $\kappa$ that characterizes a given jet \cite{Abreu}.

\parindent 3mm
 For large values of $x\sim1$ the ratio of the logarithmic
 derivatives is predicted to be close to that of the gluon and quark
 ``colour charges'', $C_A/C_F=9/4$. Experimentally, the ratio was
 measured to be
\end{minipage}
\begin{equation}
  \label{eq:cacf}
\frac{C_A}{C_F}= 2.23  \pm 0.09_{\mbox{\scriptsize stat.}} \pm 
0.06_{\mbox{\scriptsize syst.}}.    
\end{equation}

\subsection{Bremsstrahlung parton vs.\ hadron multiplicities (global
            direct evidence)}

\noindent
\begin{minipage}{0.42\textwidth}
\parindent 3mm
 Since accompanying QCD radiation seems to be there, we can make a
 step forward by asking for a {\em direct}\/ evidence: what is the
 fate of those gluons and sea quark pairs produced via multiple
 initial gluon bremsstrahlung followed by parton multiplication
 cascades?  

\parindent 3mm
 Let us look at the $Q$-dependence of the mean hadron multiplicity,
 the quantity dominated by relatively soft particles with
 $x\ll1$. This is the kinematical region populated by accompanying QCD
 radiation.

\parindent 3mm
 Fig.~6  demonstrates that the hadron multiplicity increases with the
 hardness of the jet proportional to the multiplicity of secondary
 gluons and sea quarks.
\end{minipage}
\hfill
\begin{minipage}{0.55\textwidth}
\epsfig{file=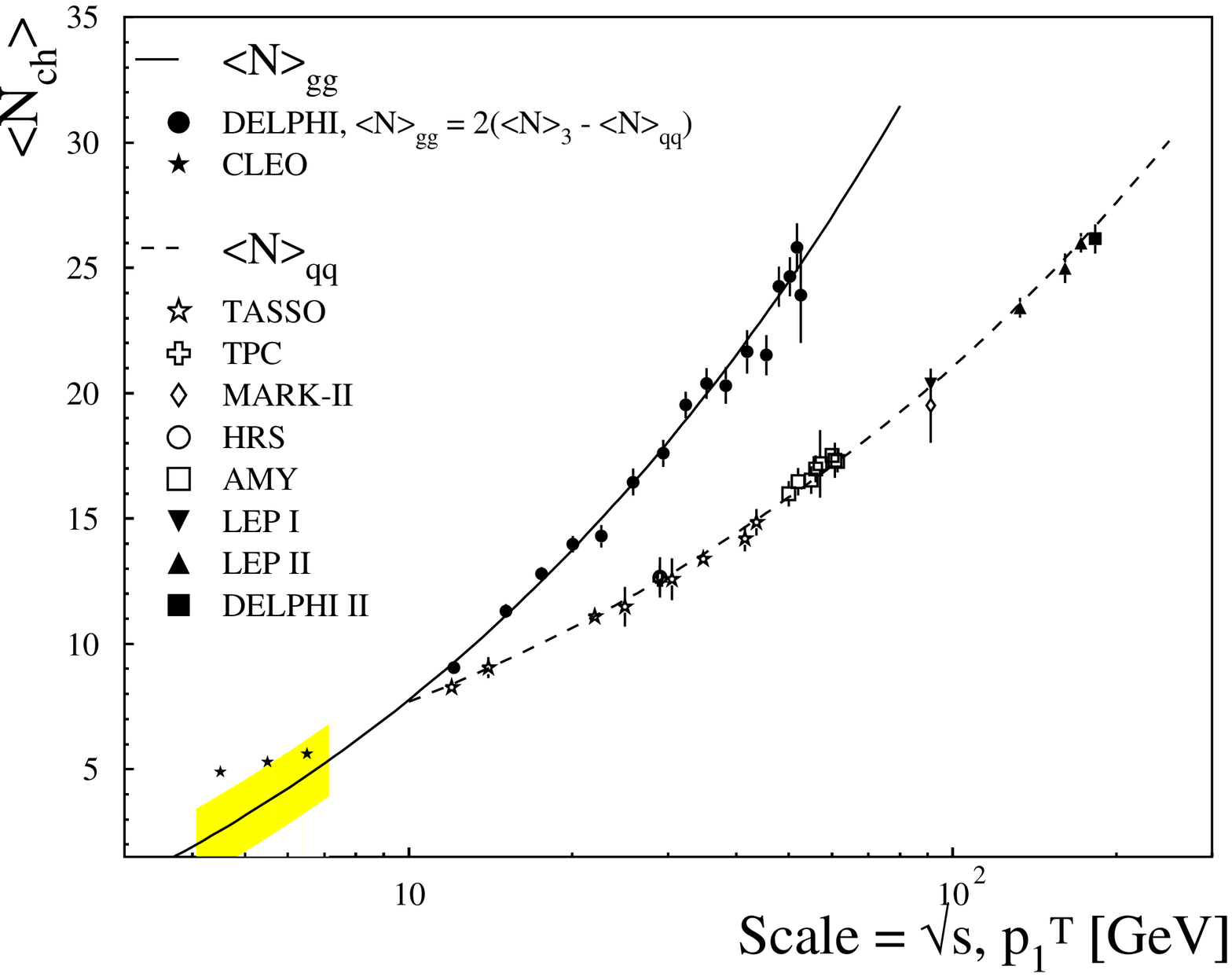,width=\textwidth}

Fig.~6: DELPHI comparison of charged hadron multiplicities from tagged
quark and gluon jets {\protect\cite{Abreu}}.
\end{minipage}

\noindent
The ratio of the slopes, once again, provides an independent measure
of the ratio of the colour charges, which is consistent with
\eqref{eq:cacf} \cite{Abreu}:
\begin{equation}
\frac{C_A}{C_F}= 2.246  \pm 0.062_{\mbox{\scriptsize stat.}} \pm 
0.008_{\mbox{\scriptsize syst.}} \pm 0.095_{\mbox{\scriptsize theo.}}.      
\end{equation}

\vspace{3mm}

 Since the total numbers match, it is time to ask a more delicate
 question about energy-momentum distribution of final hadrons versus
 that of the underlying parton ensemble.  One should not be too picky
 in addressing such a question.  It is clear that hadron-hadron
 correlations, for example, will show resonant structures about which
 the quark-gluon speaking pQCD can say little, if anything, at the
 present state of the art.  Inclusive single-particle distributions,
 however, have a better chance to be closely related.  Triggering a
 single hadron in the detector, and a single parton on paper, one may
 compare the structure of the two distributions to learn about
 dynamics of hadronization.

 It is important to stress that QCD coherence is crucial for treating
 particle multiplication {\bf inside} jets, as well as for hadron
 flows {\bf in-between} jets.

\subsection{Multiplicity flows between jets (another global
            direct tricky evidence)}

 ``QCD Radiophysics'' deals with particle flows in the angular regions
 {\em between jets}\/ in various multi-jet configurations.  These
 particles do not belong to any particular jet, and their production,
 at the pQCD level, is governed by {\em coherent}\/ soft gluon
 radiation off the multi-jet system as a whole as off a composite {\em
 antenna}\/ (hence, ``radiophysics'').

 The \underline{ratios} of particle (gluon) flows in different
 inter-jet valleys are given by parameter-free \PT\1 predictions and
 reveal the so-called ``string''~\cite{Lund_string} or ``drag''
 effects~\cite{drag}.

 At the level of the \PT\ accompanying gluon radiation ({\em QCD
 radiophysics}\/) such ratios are quite simple and straightforward to
 derive.  They depend only on the number of colours ($N_c$) and on the
 geometry of the underlying ensemble of hard partons forming jets.

\paragraph{Lund string effect.}
 For example, the classical {\em string effect}\/ -- the ratio of the
 multiplicity flow between a quark (antiquark) and a gluon to that in
 the $ q\bar{q}$ valley in symmetric (``Mercedes'') $q\bar{q}g$
 three-jet $e^+e^-$ annihilation events reads
$$
 \frac{dN^{(q\bar{q}g)}_{{ qg}}}{dN^{(q\bar{q}g)}_{{ q\bar{q}}}} 
  \simeq  \frac{5N_c^2-1}{2N_c^2-4}= \frac{22}{7} \>\simeq\>\pi. 
$$
 We see that emitting an energetic gluon off the initial quark pair
 depletes accompanying radiation in the {\em backward}\/ direction:
 colour is {\em dragged out}\/ of the $ q\bar{q}$ valley. This
 destructive interference effect is so strong that the resulting
 multiplicity flow between quarks falls below that in the {\em least
 favourable}\/ direction {\em transversal}\/ to the 3-jet event plane:
$$
  \frac{dN^{(q\bar{q}\gamma)}_{\perp}} {dN^{(q\bar{q}g)}_{q\bar{q}}}  
  \simeq \frac{N_C+2C_F}{2(4C_F-N_c)}= \frac{17}{14}\,.
$$
The following pictures demonstrate the DELPHI study of the particle
flow in the out-of-plane direction, as a function of the angle
$\Theta_1$ between the two softer jets (one of them the
gluon)~\cite{Ham}.  The particle flow increases with angle in a full
accord with the theoretical expectation based on the coherent gluon
radiation off the three-prong colour antenna.


\begin{minipage}{0.4\textwidth}
\epsfig{file=d_tamp2.eps,width=0.95\textwidth,height=1.3\textwidth}
%
\end{minipage}
\begin{minipage}{0.6\textwidth}
\epsfig{file=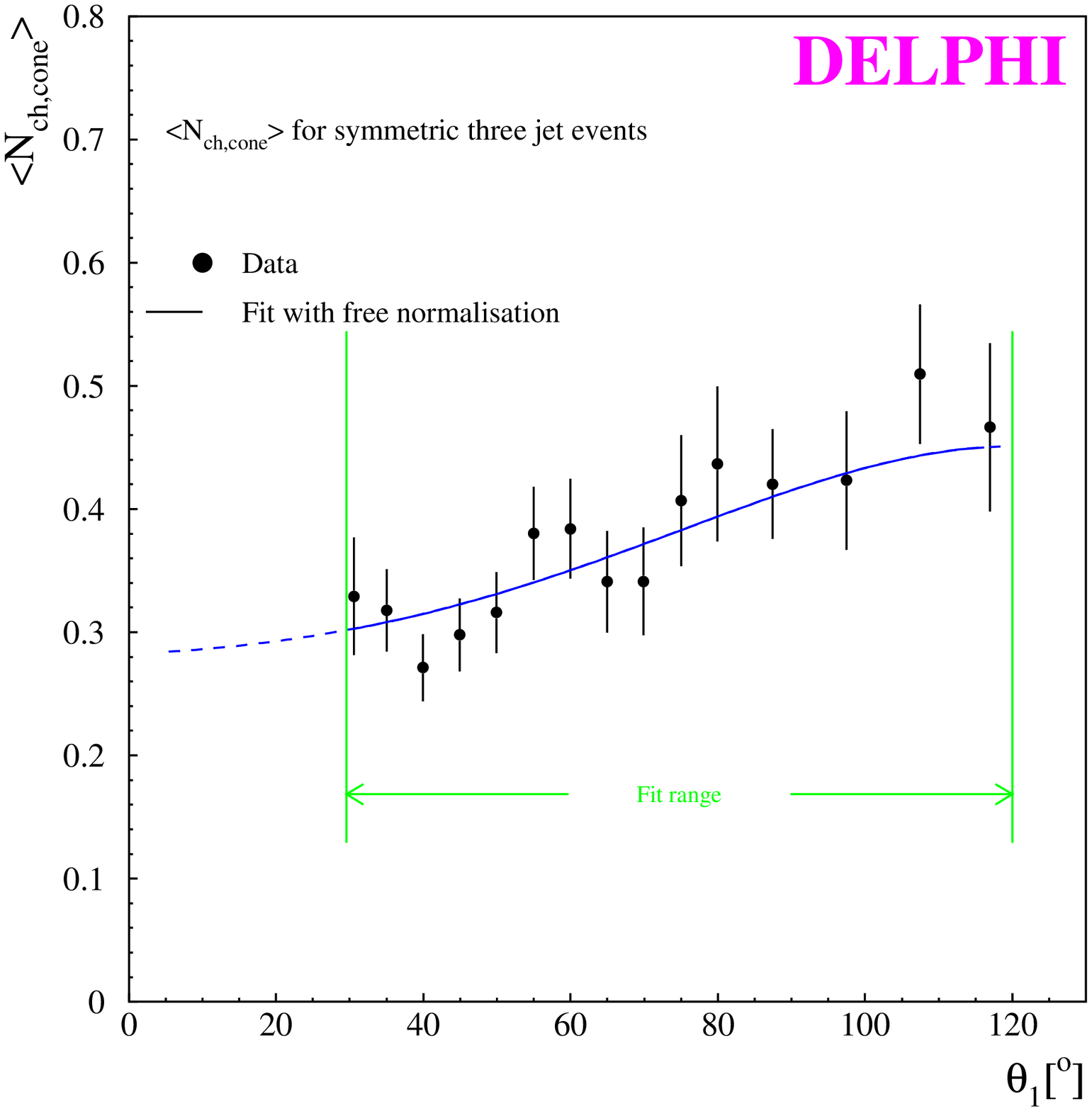,width=0.95\textwidth}
\end{minipage}


\paragraph{Another Example.}
 A comparison of the hadron flows in the $q\bar{q}$ valley in $
 q\bar{q}\gamma$ events with a gluon jet replaced by an energetic
 photon results in the ratio
$$
\frac{dN^{(q\bar{q}{{\gamma}})}_{q\bar{q}}}{dN^{(q\bar{q}{{g}})}_{q\bar{q}}}
\simeq \frac{2(N_c^2-1)}{N_c^2-2}= \frac{16}{7}, 
\quad \mbox{({experiment}: $ 2.3\pm 0.2$).}  
$$
It is not strange at all that with {\em gluons}\/ one can get, e.g.,
\begin{equation}\label{quantum}
\qq:\quad  1+1\>=\>2\qquad \mbox{while} \qquad  
\qq+g:\quad  1+ 1\>+\> \frac{9}{4} \>=\>
 \frac{7}{16},
\end{equation}
 which is simply the radiophysics of composite antennas, or {\em
 quantum mechanics}\/ of conserved colour charges.  The first equation
 of these quantum arithmetics problems describes symbolically the
 density of soft gluon radiation between two quarks in a
 $q\bar{q}\gamma$ event, with 1 standing for the colour quark charge.

 Replacing the colour-neutral photon by a gluon one gets an additional
 emitter with the relative strength $\frac94$, as shown in the l.h.s.\
 of the second equation in~\eqref{quantum}. In spite of having added
 an additional emitter, the resulting soft gluon yield in the
 $q\bar{q}$ direction (r.h.s.)  {\em decreases}\/ substantially as a
 result of destructive interference between three elements of a
 composite colour antenna.

 Nothing particularly strange, you might say.  What {\em is}\/ rather
 strange, though, is that this naive perturbative\1 wisdom is being
 impressed upon junky 100--300 \MeV\ pions which dominate hadron flows
 between jets in the present-day experiments such as the OPAL study
 shown in Fig.~7.

\noindent
\begin{center}
\begin{minipage}{0.75\textwidth}
\epsfig{figure=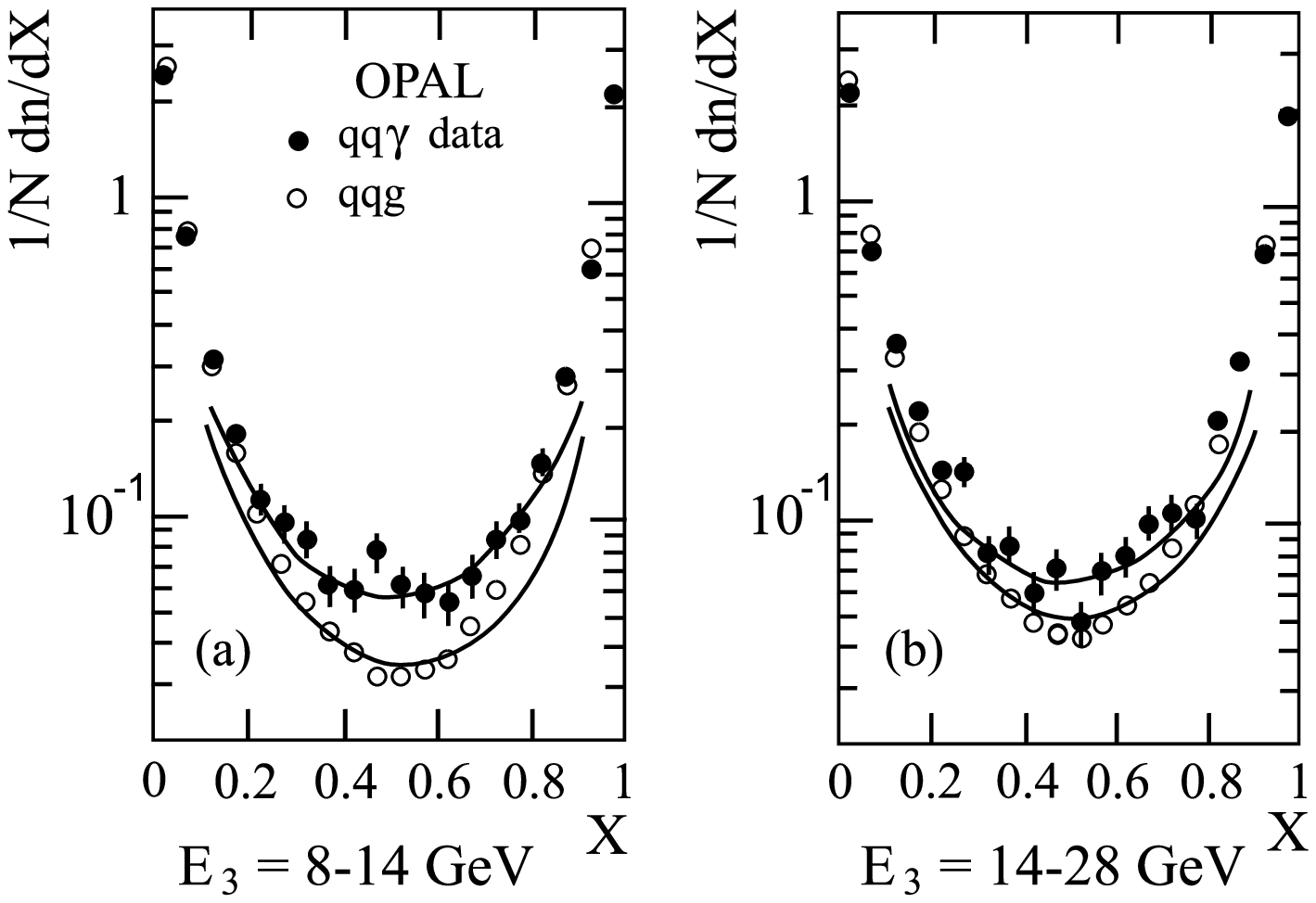,width=\textwidth} 

\vspace{-8cm}

\noindent
Fig.7: Particle flows in the $q\bar{q}$ valley in $q\bar{q}\gamma$ and
$q\bar{q}g$ events~\cite{OPAL_drag} versus an analytic parameter-free
prediction based on the soft gluon radiation pattern~\cite{Buican}.
\end{minipage}
\end{center}



\vspace {3mm}

 These and many similar phenomena are being seen experimentally.  What
 the nature seems to be telling us, is that
\begin{itemize}
\item The {\em colour field}\/ that an ensemble of hard primary {\bf
    partons} (parton antenna) develops, determines, on the one-to-one
  basis, the structure of final flows of {\bf hadrons}.
\item The Poynting vector of the colour field gets translated into the
  hadron Poynting vector without any visible reshuffling of particle
  momenta at the hadronization stage.
\end{itemize}
 When viewed {\em globally}, confinement is about {\em renaming}\/ a
 flying-away quark into a flying-away pion rather than about forces
 {\em pulling}\/ quarks together.

\subsection{Inclusive hadron distribution inside jets (local direct evidence)}

 A similar message comes from the study of the energy distribution of
 particles {\em inside}\/ jets.

 An inclusive energy spectrum of soft bremsstrahlung partons in QCD
 jets has been derived in 1984 in the so-called MLLA -- the Modified
 Leading Logarithmic Approximation~\cite{LPHD,Mueller83}.  This
 approximation takes into account all essential ingredients of parton
 multiplication in the next-to-leading order. They are: parton
 splitting functions responsible for the energy balance in parton
 splitting, the running coupling $\alpha_s(k_\perp^2)$ depending on
 the relative transverse momentum of the two offspring and exact
 angular ordering.
 The latter is a consequence of soft gluon coherence and plays, as we
 shall discuss below, an essential r\^ole in parton dynamics.  In
 particular, gluon coherence suppresses multiple production of very
 small momentum gluons. It is particles with intermediate energies
 that multiply most efficiently.  As a result, the energy spectrum of
 relatively soft secondary partons in jets acquires a characteristic
 hump-backed shape.
 The position of the maximum in the logarithmic variable $\xi=-\ln x$,
 the width of the hump and its height increase with $Q^2$ in a
 predictable way.

 The shape of the inclusive spectrum of all charged hadrons (dominated
 by $\pi^{\pm}$) exhibits the same features. This comparison,
 pioneered by Glen Cowan (ALEPH) and the OPAL collaboration, has since
 become a standard test of analytic QCD predictions.
First scrutinized at LEP, the similarity of parton and hadron energy
distributions has been verified at SLC and KEK $e^+e^-$ machines, as
well as at HERA and Tevatron where hadron jets originate not from bare
quarks dug up from the vacuum by a highly virtual photon/$Z^0$ but
from hard partons kicked out from initial hadron(s).

 In Fig.~8 (DELPHI) the comparison is made of the spectra of all
 charged hadrons at various annihilation energies $Q$ with the
 so-called ``distorted Gaussian'' fit~\cite{FW89}
 which employs the first four moments 
\noindent
\begin{minipage}{0.5\textwidth}
\begin{center}
\epsfig{file=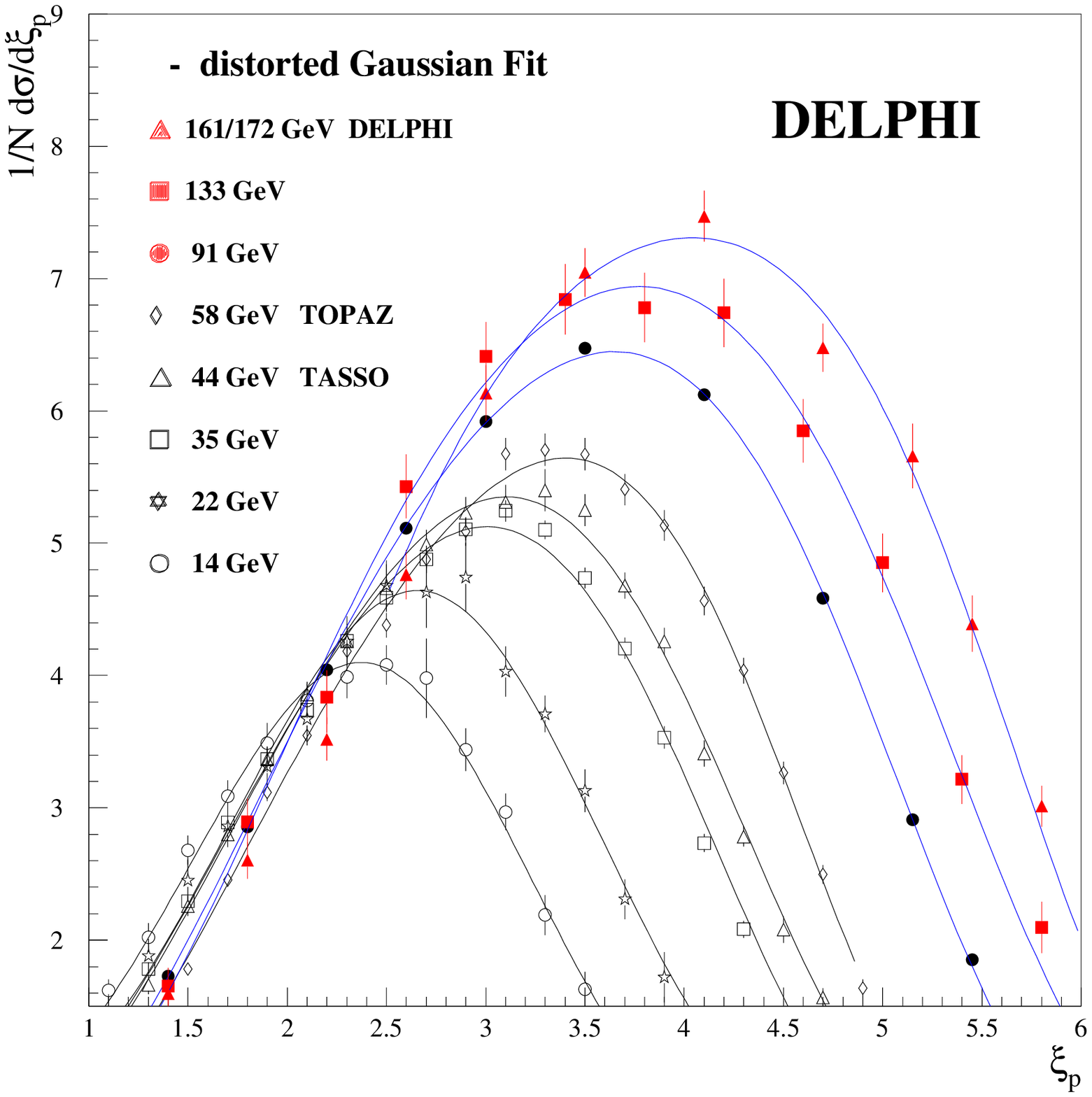,width=\textwidth}

Fig.~8: Hadron energy spectra in $e^+e^-\to h^{\pm}+X$
\end{center}
\end{minipage}
\hfill
\begin{minipage}{0.45\textwidth}
(the mean, width, skewness and kurtosis) of the MLLA distribution
around its maximum.

\parindent 3mm
 Shall we say, a (routine, interesting, wonderful) check of yet
 another QCD prediction?  

\parindent 3mm 
 I would rather not. Such a close similarity offers a deep puzzle,
 even a worry, rather than a successful test.

\parindent 3mm
 Indeed, after a little exercise in translating the values of the
 logarithmic variable $\xi=\ln(E_{\mbox{\scriptsize jet}}/p)$ in
 Fig.~8  into {\GeV}s you will see that the actual hadron momenta at
 the maxima are, for example, $p$=$\half Q\cdot
 e^{-\xi_{\max}}\simeq$~0.42, 0.85 and 1.0~\GeV\ for $Q$=14, 35~\GeV\
 and at LEP-I, $Q$=91~\GeV.  Is it not surprising that the \PT\
 spectrum is mirrored by that of the pions (which constitute 90\%\ of
 all charged hadrons produced in jets) with momenta well below
 1~{\GeV}?!

\end{minipage}

\vspace{2mm}

 For this very reason the observation of the parton-hadron similarity
 was initially met with a serious and well grounded scepticism: it
 looked more natural (and was more comfortable) to blame the finite
 hadron mass effects for falloff of the spectrum at large $\xi$ (small
 momenta) rather than seriously believe in applicability of the \PT\1\2
 consideration down to such disturbingly small momentum scales.

 This worry has been answered by CDF. Andrey Korytov and Alexei
 Safonov carried out meticulous studies of the energy distribution of
 hadrons produced inside a restricted cone of the opening half-angle
 $\Theta_c$ around the jet axis.

 As we have already mentioned above discussing the Lund
 hadroproduction picture, theoretically it is not the energy of the
 jet but the maximal parton transverse momentum inside it,
 $k_{\perp\max}\simeq E_{\mbox{\scriptsize jet}} \sin\Theta_c$, that
 determines the hardness scale and thus the yield and the distribution
 of the accompanying radiation~\cite{cone}.
 This means that by choosing a small opening angle one can study
 relatively small hardness scales but in a cleaner environment: due to
 the Lorentz boost effect, eventually all particles that form a short
 small-$Q^2$ QCD ``hump'' are now relativistic and concentrated at the
 tip of the jet.
 For example, selecting hadrons inside a cone $\Theta_c\simeq 0.14$
 around an energetic quark jet with $E_{\mbox{\scriptsize jet}}\simeq
 100$~\GeV\ (LEP-II) one would see that very ``dubious'' $Q=14$~\GeV\
 curve in Fig.~8 but now with the maximum boosted from 0.45~\GeV\ into
 a comfortable 6 \GeV\ range.

 The CDF Fig.~9 \cite{Goulianos,Safonov} shows the change of the
 energy spectrum of charged hadrons with the opening angle for a given
 invariant mass of the system of two large transverse momentum jets,
 in comparison with the analytic MLLA expression for soft secondary
 gluons. Similar results for a broad range of dijet masses, $78\,\GeV
 \le M_{jj} \le 537\,\GeV$, will soon be made public.
%
 A close similarity between the hadron yield and the full MLLA parton
 spectra can no longer be considered accidental and be attributed to
 non-relativistic kinematical effects.

\subsection{Brave gluon counting}

 Modulo $\Lambda_{\mbox{\scriptsize QCD}}$, there is only one unknown
 in this comparison, namely, the overall normalization of the spectrum
 of hadrons relative to that of partons (bremsstrahlung gluons).

 Strictly speaking, there should/could have been another free
 parameter, that which quantifies one's bravery in applying the pQCD
 dynamics. It is the minimal transverse momentum cutoff in parton
 cascades, $k_\perp>Q_0$.  The strength of successive $1\to2$ parton
 splittings is proportional to $\alpha_s(k_\perp^2)$ and grows with
 $k_\perp$ decreasing.  The necessity to terminate the process at some
 low transverse momentum scale where the \PT\ coupling becomes large
 (and eventually hits the formal ``Landau pole'' at
 $k_\perp=\Lambda_{\mbox{\scriptsize QCD}}$) seems imminent.
 Surprisingly enough, it is not.

\begin{center}
\epsfig{file=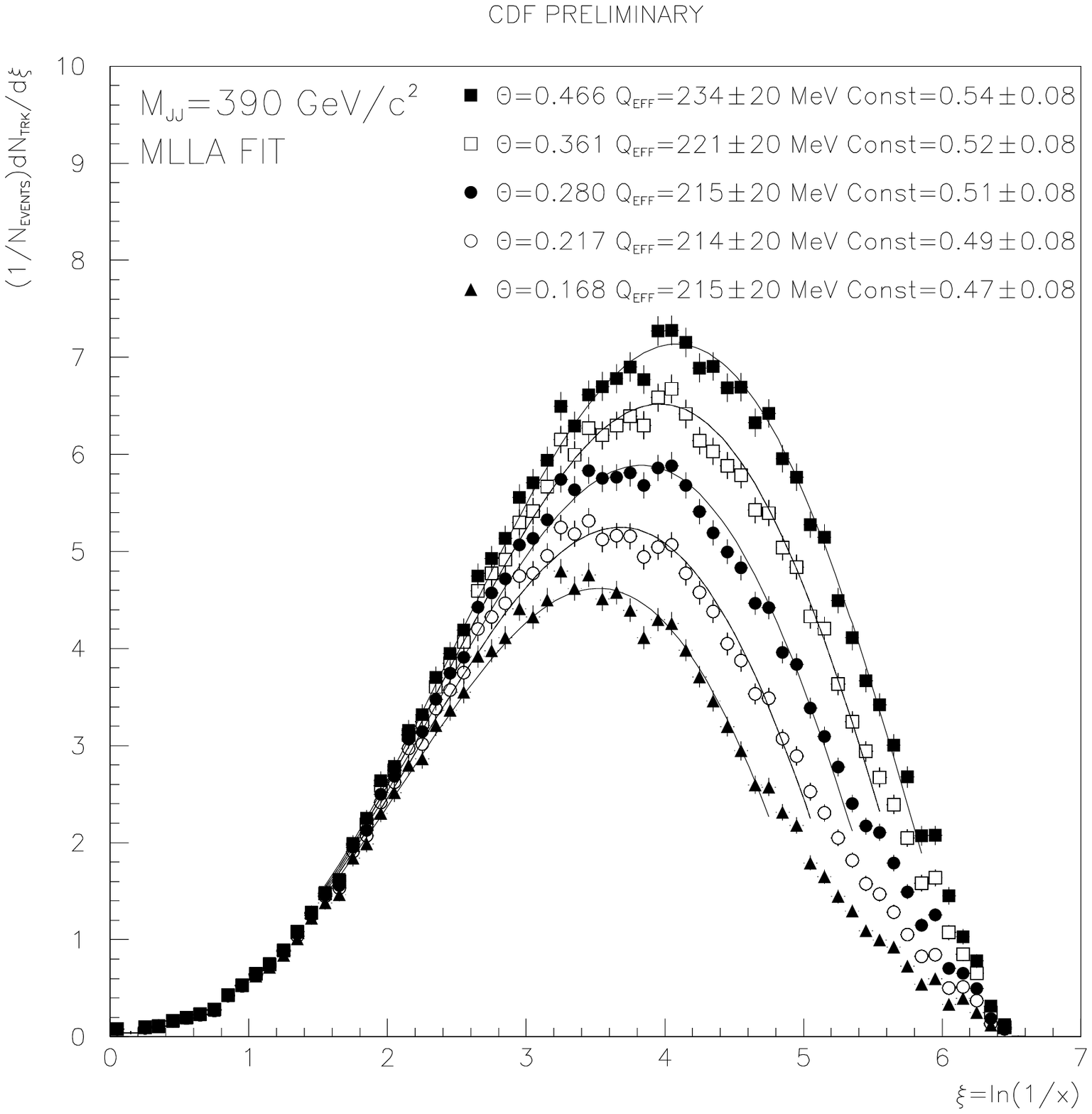,width=0.7\textwidth}

Fig.~9: {Inclusive energy distributions of charged hadrons in 
large--$p_\perp$ jets {\protect\cite{Goulianos}}.
}
\end{center}

 As we shall see in the next Section, the inclusive parton energy
 distribution turns out to be a CIS QCD prediction, believe it or not.
 It is its crazy $Q_0=\Lambda_{\mbox{\scriptsize QCD}}$ limit (the
 so-called ``limiting spectrum'') which is shown by solid curves in
 Fig.~9.

 Choosing the minimal value for the collinear parton cutoff $Q_0$ can
 be looked upon as shifting, as far as possible, responsibility for
 particle multiplication in jets to the \PT\ dynamics.  This brave
 choice can be said to be dictated by experiment, in a certain
 sense. Indeed, with increase of $Q_0$ the parton distributions {\em
 stiffen}\/ (parton energies are limited from below by the kinematical
 inequality $xE_{\mbox{\scriptsize jet}}\equiv k \ge k_\perp > Q_0$).
 The maxima would move to larger $x$ (smaller $\xi$), departing from
 the data.

\noindent
\begin{minipage}{0.55\textwidth}
\begin{center}
\epsfig{file=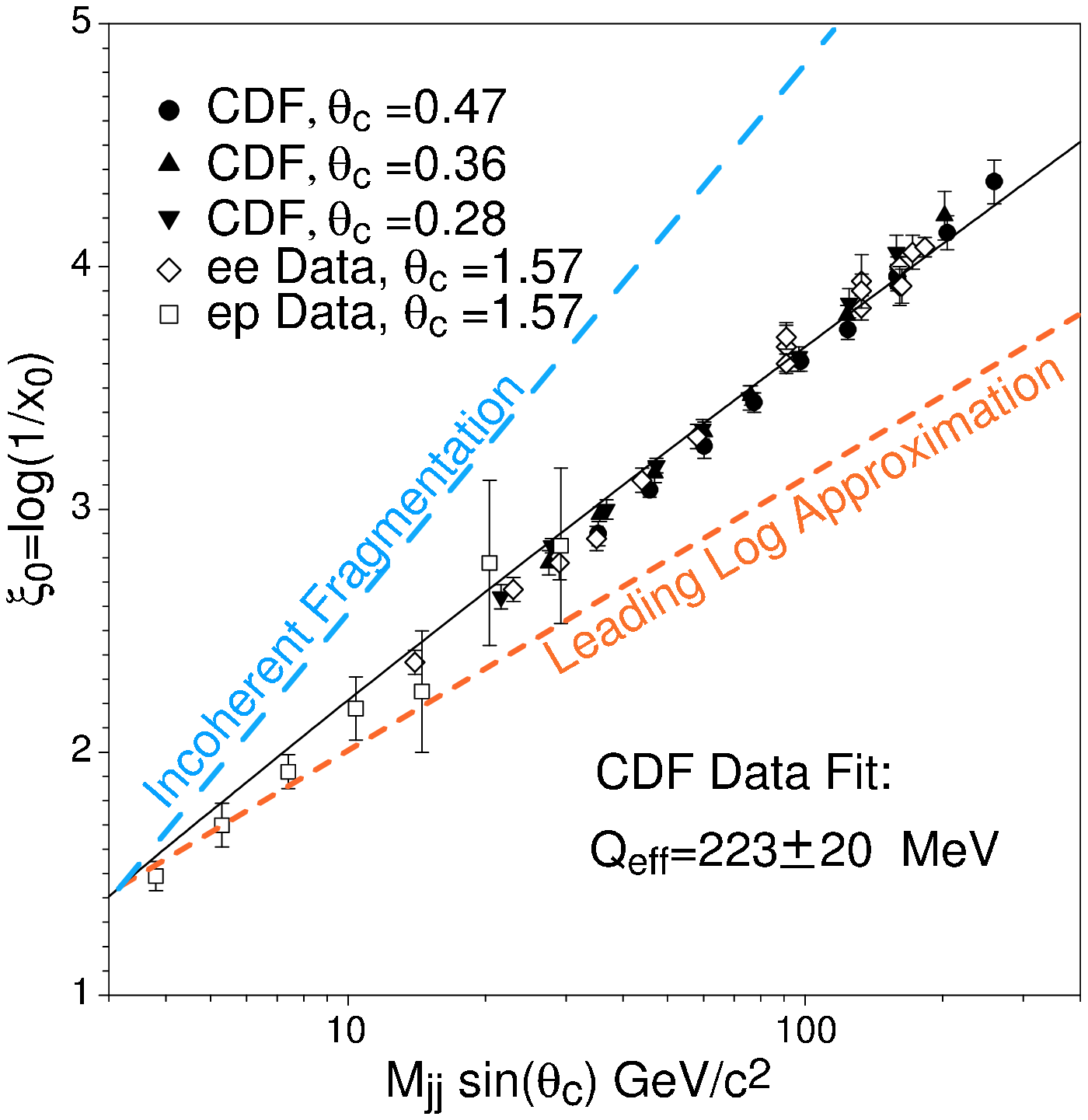,width=\textwidth}

Fig.~10: {Position of the maximum versus MLLA~\cite{Safonov_New}.}
\end{center}
\end{minipage}
\hfill
\begin{minipage}{0.4\textwidth}
 A clean test of ``brave gluon counting'' is provided by Fig.~10 where
 the position of the hump, which is insensitive to the overall
 normalization, is compared with the parameter-free analytic MLLA
 prediction~\cite{Safonov_New}. An overlaid prediction of the
 incoherent hadronization model (long-dashed line) shows how the
 maximum of the energy distribution would have moved if the
 Field--Feynman fragmentation picture were applicable. Comparison with
 the DLA expectation (short dashes) demonstrates the r\^ole of the NLO
 effects parton cascades (MLLA).

\parindent 3mm
 Spectacular verification of the local duality hypothesis was
 recently reported by A.~Safonov~\cite{Safonov_New}. It showed
 remarkable stability of the only parameter of the game --
 $Q_{\mbox{\scriptsize eff}}$ --
 wrt variations of
 the dijet mass $M_{JJ}$ and the opening angle of the cone $\Theta_c$
 (left template in Fig.~11).
\end{minipage}


 This parameter plays a double r\^ole in the naive limiting spectrum:
 that of $\LQCD$ and of the collinear cut $Q_0$.  The evolution of the
 spectrum with the hardness of the process ($M_{jj}\sin\Theta_c$)
 obviously depends on $\LQCD$, while the position of the maximum is
 sensitive to $Q_0$. The right template in Fig.~11 demonstrates an
 impressive correlation between the two independent determinations of
 $Q_{\mbox{\scriptsize eff}}$~\cite{Safonov_New}.


\noindent
\begin{center}
\begin{minipage}{0.48\textwidth}
\epsfig{file=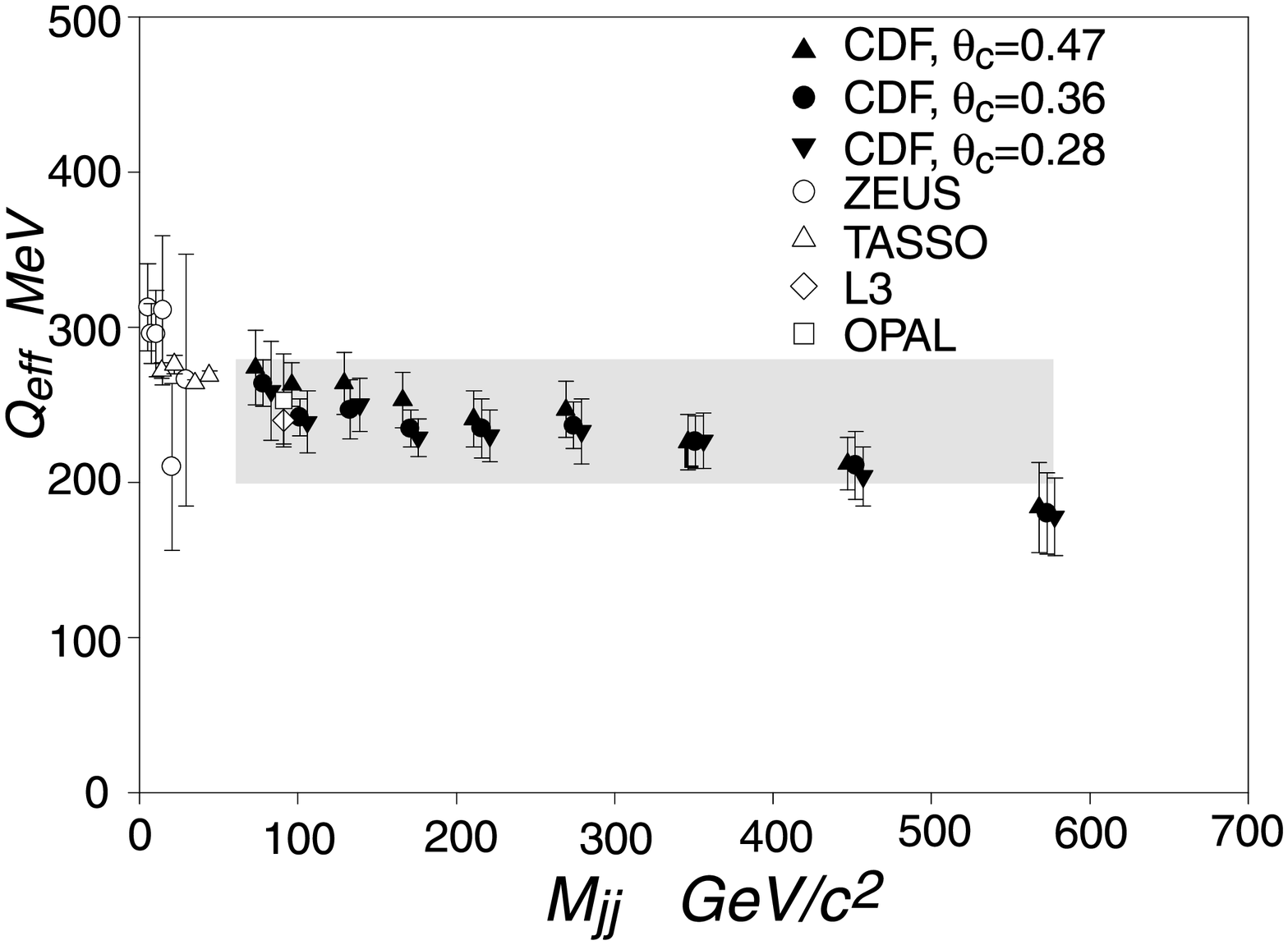,width=\textwidth,height=0.9\textwidth}
\end{minipage}
\hfill
\begin{minipage}{0.48\textwidth}
\epsfig{file=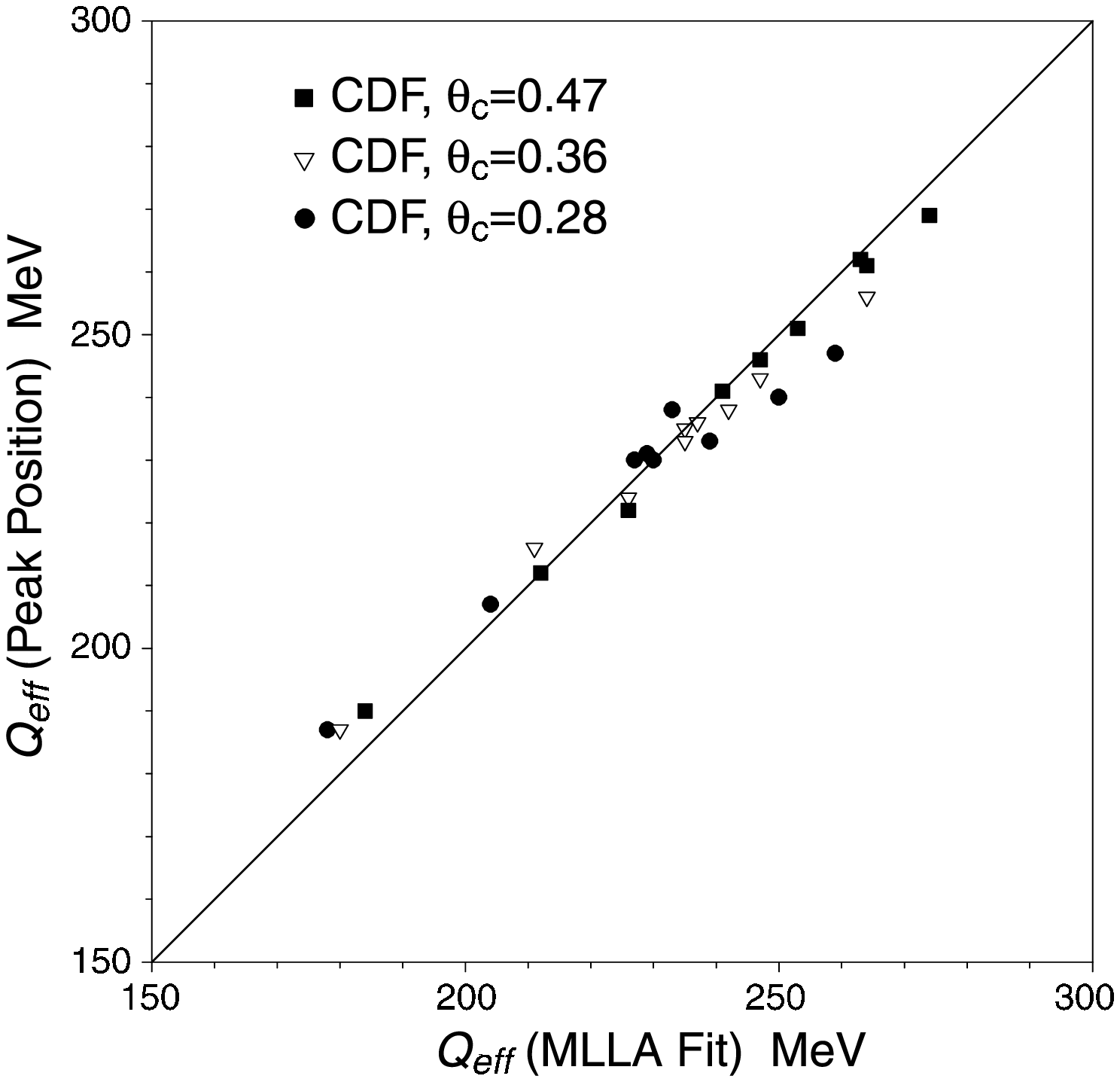,width=\textwidth,height=0.9\textwidth}
\end{minipage}
Fig.~11: Experimental results on the $Q_{\mbox{\scriptsize eff}}$
parameter of the limiting MLLA spectrum~\cite{Safonov_New}.

\end{center}

 \section{HUMPBACKED PLATEAU AND THE ORIGIN OF LPHD}

 Here we are going to derive together the QCD ``prediction'' of the
 inclusive energy spectrum of relatively soft particles from QCD jets.
 I put the word {\em prediction}\/ in quotation marks on purpose.
 This is a good example to illustrate the problem of filling the gap
 between the QCD formulae, talking quarks and gluons, and phenomena
 dealing, obviously, with hadrons.

 Let me first make a statement:
\begin{quote}
 It is QCD coherence that allows the prediction of the inclusive soft
 particle yield in jets practically from ``first principles''.
\end{quote}
 You have all the reasons to feel suspicious about this.  Indeed, we
 have stressed above the similarity between the dynamics of the
 evolution of space-like (DIS structure functions) and time-like
 systems (jets).  On the other hand, you are definitely aware of the
 fact that the DIS structure functions cannot be calculated
 perturbatively.

 In spite of the similarity between the space- and time-like evolution
 of {\em hard}\/ partons, $x\sim1$, there is an essential difference
 between {\em small}--$x$ physics of DIS structure functions and the
 jet fragmentation.
 In the case of the {\em space-like evolution}, in the limit of small
 Bjorken--$x$ the problem becomes essentially non-perturbative and
 pQCD loses control of the DIS cross sections~\cite{CC}. In contrast,
 studying small-Feynman-$x$ particles originating from the {\em
 time-like evolution}\/ of jets offers a gift and a puzzle: all the
 richness of the confinement dynamics reduces to a mere overall
 normalization constant.

\subsection{Solving the DIS evolution}

 So let us repeat that DIS structure functions at $x\sim 1$ cannot be
 calculated perturbatively\1 from first principles. Indeed there are
 input parton distributions for the target proton, which have to be
 plugged in as an initial condition for the evolution at some finite
 hardness scale $Q_0=\cO{1\,\GeV}$.
 These initial distributions cannot be calculated ``from first
 principles'' nowadays but are subject to fitting.  What pQCD controls
 then, is the scaling violation pattern. Namely, it tells us how the
 parton densities change with the changing scale of the
 transverse-momentum probe:
\beq\label{eveqD} 
 \frac{\partial}{\partial \ln k_\perp}
 D(x,k_\perp) = \frac{\as(k_\perp)}{\pi} \int_x^1\frac{d z}{z}\>
 P(z)\, D\left(\frac{x}{z},k_\perp\right) .  
\eeq 
 It is convenient to present our ``wavefunction'' $D$ and
 ``Hamiltonian'' $P$ in terms of the complex moment $\omega$, which is
 Mellin conjugate to the momentum fraction $x$:
\begin{subequations}
\label{Mellin}
\beeq
D_\omega = \int_0^1 d x\>x^{\omega} \cdot D(x)\,, &&
D(x) = x^{-1}\int_{(\Gamma)} \frac{d\omega}{2\pi \I}\> 
x^{-\omega} \cdot D_{\omega}\,;\qquad { }\\
P_\omega = \int_0^1 d z\>z^{\omega} \cdot P(z)\,, &&
P(z) = z^{-1}\int_{(\Gamma)} \frac{d\omega}{2\pi \I}\> 
z^{-\omega} \cdot P_{\omega}\,,
%
\eeeq
\end{subequations}
 where the contour $\Gamma$ runs parallel to the imaginary axis, to
 the right from singularities of $D_\omega$ ($P_\omega$).  It is like
 trading the coordinate ($\ln x$) for the momentum ($\omega$) in a
 Schr\"odinger equation.

 Substituting (\ref{Mellin}) into (\ref{eveqD}) we see that the
 evolution equation becomes algebraic and describes propagation in
 ``time'' $d t=\alpi d\ln k_\perp$ of a free quantum mechanical
 ``particle'' with momentum $\omega$ and the dispersion law
 $E(\omega)=P_\omega$:
\beq\label{Schr} 
 \d \> D_\omega(k_\perp) \>=\>\frac{\as(k_\perp)}{\pi} \cdot P_\omega\>
 D_\omega(k_\perp)\,; \qquad \d\equiv \frac{\partial}{\partial \ln
 k_\perp}\>.
\eeq 
 To continue the analogy, our wavefunction $D$ is in fact a
 multi-component object. It embodies the distributions of valence
 quarks, gluons and secondary sea quarks which evolve and mix
 according the $2\times 2$ matrix ``Hamiltonian'' of the parton
 splitting functions $P[A\to B]$.

 At small $x$, however, the picture simplifies. Here the valence
 distribution is negligible, $\cO{x}$, while the gluon and sea quark
 components form a system of two coupled oscillators which is easy to
 diagonalize.
 What matters is one of the two energy eigenvalues (one of the two
 branches of the dispersion rule) that is {\em singular}\/ at
 $\omega=0$.  The problem becomes essentially one-dimensional. Sea
 quarks are driven by the gluon distribution while the latter is
 dominated by gluon cascades.  Correspondingly, the leading energy
 branch is determined by gluon-gluon splitting $g\to gg$, with a
 subleading correction coming from the $g\to q(\bar{q})\to g$
 transitions,
\beq\label{Plead}
  P_\omega \>=\> \frac{2N_c}{\omega} -a \>+\> \cO{\omega}\>, \quad
  a = \frac{11N_c}{6}+ \frac{n_f}{3N_c^2}\,.
\eeq

 The solution of (\ref{Schr}) is straightforward:
\begin{subequations}
\beeq\label{yd_sfandim}
\label{Dexp}
 D_\omega(k_\perp) &=& D_\omega(Q_0)\cdot \exp\left\{
 \int^{k_\perp}_{Q_0} \frac{dk}{k} \> \gamma_\omega(\as(k))\right\},
 \\
 \label{gamma}
 \gamma_\omega(\as) &=& \frac{\as}{\pi} \, P_\omega\,.  
\eeeq 
\end{subequations}
 The structure (\ref{Dexp}) is of the most general nature.  It follows
 from {\em renormalizability}\/ of the theory, and does not rely on
 the LLA which we used to derive it.  The function $\gamma(\as)$ is
 known as the ``anomalous dimension''\footnote{The name is a relict
 of those good old days when particle and solid state physicists used
 to have common theory seminars. If the coupling $\as$ were constant
 (had a ``fixed point''), then (\ref{Dexp}) would produce the function
 with a non-integer (non-canonical) dimension $D(Q)\propto Q^{\gamma}$
 (analogy -- critical indices of thermodynamical functions near the
 phase transition point).}.
 It can be perfected by including higher orders of the \PT\
 expansion. Actually, modern analyses of scaling violation are based
 on the improved next-to-LLA (two-loop) anomalous dimension, which
 includes $\as^2$ corrections to the LLA expression (\ref{gamma}).

 The structure (\ref{Dexp}) of the $x$-moments of parton distributions
 (DIS structure functions) gives an example of a clever separation of
\PT\ and \NP\ effects; in this particular case -- in the form of two 
factors. 
 It is the $\omega$-dependence of the input function $D_\omega(Q_0)$
 (``initial parton distributions'') that limits predictability of the
 Bjorken-$x$ dependence of DIS cross sections.

 So, how comes then that in the time-like channel the \PT\ answer
 turns out to be more robust?

\subsection{Coherent hump in $e^+e^-\to h(x) + \ldots$} 

 We are ready to discuss the time-like case, with $D^h_j(x,Q)$ now the
 inclusive distribution of particles $h$ with the energy fraction
 (Feynman-$x$) $x\ll1$ from a jet (parton $j$) produced at a large
 hardness scale~$Q$.
 
 Here the general structure (\ref{Dexp}) still holds. We need,
 however, to revisit the expression (\ref{gamma}) for the anomalous
 dimension because, as we have learned, the proper evolution time is
 now different from the case of DIS.

 In the time-like jet evolution, due to Angular Ordering, the
 evolution equation becomes non-local in $k_\perp$ space:
\beq\label{eveqE} 
\frac{\partial}{\partial \ln k_\perp} D(x,k_\perp) = \frac{\as(k_\perp)}{\pi}
\int_x^1 \frac{d z}{z}\> P(z)\, D\left(\frac{x}{z},\, z\cdot k_\perp\right).
\eeq 
Indeed, successive parton splittings are ordered according to 
$$
 \theta = \frac{k_\perp}{k_{||}} \> > \>    
 \theta'= \frac{k'_\perp}{k'_{||}} \,. 
$$
Differentiating $D(k_\perp)$ over the scale of the ``probe'',
$k_\perp$, results then in the substitution
$$
    k'_\perp = \frac{k'_{||}}{k_{||}}\cdot k_\perp \>\equiv\> z\cdot k_\perp
$$ 
in the argument of the distribution of the next generation $D(k'_\perp)$.

The evolution equation (\ref{eveqE}) can be elegantly cracked 
using the Taylor-expansion trick,
\beq\label{trick}
  D(z\cdot k_\perp) =  \exp\left\{ \ln z \frac{\partial}{\partial \ln
      k_\perp} \right\}  D(k_\perp)\>=\>
z^{\frac{\partial}{\partial \ln  k_\perp}} \cdot D(k_\perp).
\eeq
Turning as before to moment space (\ref{Mellin}), we observe that the
solution comes out similar to that for DIS, (\ref{yd_sfandim}), but
for one detail. The exponent $\d$ of the additional $z$-factor in
(\ref{trick}) combines with the Mellin moment $\omega$ to make the
argument of the splitting function $P$ a {\em differential operator}\/
rather than a complex number:
\beq 
\d\cdot D_\omega = \frac{\as}{\pi} \, P_{\omega+\d} \cdot D_\omega\,.
\eeq 
This leads to the differential equation 
\beq\label{yd_difeq} 
\left(  P^{-1}_{\omega+\d}\>\d \>-\> \frac{\as}{\pi}
  -\left[P^{-1}_{\omega+\d}, \frac{\as}{\pi}\right]P_{\omega+\d}
\right) \cdot D =0\>.  
\eeq 
Recall that, since we are interested in the small-$x$ region, the
essential moments are small, $\omega\ll1$. 

For the sake of illustration, let us keep only the most singular piece
in the ``dispersion law'' (\ref{Plead}) and neglect the commutator
term in (\ref{yd_difeq}) generating a subleading correction $\propto
\d\as\sim \as^2$. In this approximation (DLA),
\beq\label{yd_andimapp} 
P_\omega \simeq \frac{2N_c}{\omega}\>,
\eeq 
(\ref{yd_difeq}) immediately gives a quadratic equation for the
anomalous dimension,\footnote{It suffices to use the next-to-leading
  approximation to the splitting function (\ref{Plead}) and to keep
  the subleading correction coming from differentiation of the running
  coupling in (\ref{yd_difeq}) to get the more accurate MLLA anomalous
  dimension~$\gamma_\omega$.
} 
\beq\label{yd_dlaeq}
(\omega+\gamma_\omega)\gamma_\omega \>-\> \frac{2N_c\as}{\pi}
+\cO{\frac{\as^2}{\omega}} \>=\>0\,.  
\eeq 

The leading anomalous dimension following from (\ref{yd_dlaeq}) is
\beq\label{yd_andimdla}
 \gamma_\omega = \frac{\omega}{2}
 \left(-1+ \sqrt{1+ \frac{8N_c\as}{\pi\, \omega^2}}\right). 
\eeq
When expanded to first order in $\as$, it coincides with that for
the space-like evolution, $\gamma_\omega\simeq\as/\pi\cdot P_\omega$, with 
$P$ given in (\ref{yd_andimapp}).
Such an expansion, however, fails when characteristic 
$\omega\sim1/|\ln x|$ becomes as small as $\sqrt{\as}$, that is when 
$$
  \frac{8N_c \as}{\pi}\, \ln^2 x \> \ga \> 1\,.
$$

Now what remains to be done is to substitute our new weird anomalous
dimension into (\ref{Dexp}) and perform the inverse Mellin transform
to find $D(x)$.  If there were no QCD parton cascading, we would expect
the particle {\em density}\/ $xD(x)$ to be constant (Feynman plateau).
It is straightforward to derive that plugging in the DLA anomalous
dimension (\ref{yd_andimdla}) results in the plateau density increasing
with $Q$ and with a maximum (hump) ``midway'' between the smallest
and the highest parton energies, namely, at $x_{\max}\simeq \sqrt{Q_0/Q}$.
The subleading MLLA effects shift the hump to smaller parton
energies, 
$$
\ln \frac{1}{x_{\max}} = \ln\frac{Q}{Q_0}\left(\frac12 +
  c\cdot\sqrt{\as}+\ldots \right) \simeq 0.65\,\ln\frac{Q}{Q_0}\,, 
$$
with $c$ a known analytically calculated number.  Moreover, defying
naive probabilistic intuition, the softest particles do not multiply
at all. The density of particles (partons) with $x\sim Q_0/Q$ stays
constant while that of their more energetic companions increases with
the hardness of the process $Q$.
 
This is a powerful legitimate consequence of pQCD coherence. 
We turn now to another, no less powerful though less legitimate,
consequence.  

\subsection{Coherent damping of the Landau singularity}
The time-like DLA anomalous dimension (\ref{yd_andimdla}), as well as
its MLLA improved version, has a curious property. Namely, in sharp
contrast with DIS, it allows the momentum integral in
(\ref{yd_sfandim}) to be extended to very small scales.  Even
integrating down to $Q_0=\Lambda$, the position of the ``Landau pole''
in the coupling, one gets a finite answer for the distribution (the
so-called {\em limiting spectrum}), simply because the $\sqrt{\as(k)}$
singularity happens to be integrable!

It would have been poor taste to trust this formal integrability,
since the very \PT\ approach to the problem (selection of dominant
contributions, parton evolution picture, etc.) relied on $\as$ being a
numerically small parameter.  However, the important thing is that,
due to time-like coherence effects, the (still perturbative but
``smallish'') scales, where $\as(k)\gg\omega^2$, contribute to
$\gamma$ basically in a $\omega$-{\em independent}\/ way, $
\gamma+{\omega}/{2} \propto \sqrt{\as(k)} \neq f(\omega) $.  This
means that ``smallish'' momentum scales $k$ affect only an overall
{\em normalization}\/ without affecting the {\em shape}\/ of the
$x$-distribution.

Since
such
 is the r\^ole of the ``smallish'' scales, it is natural to expect the
 same for the truly small -- non-perturbative -- scales where the
 partons transform into the final hadrons.  This hypothesis (LPHD)
 reduces, mathematically, to the statement (guess) that the \NP\
 factor in (\ref{Dexp}) has a finite $\omega\to0$ limit:
$$
   D^{(h)}_{\omega}(Q_0) \>\to\> K^h=\mbox{const}, \quad \omega\to0\,. 
$$

 In other words, decreasing $Q_0$ we start to lose control of the
 interaction intensity of a parton with a given $x$ and $k_\perp\sim
 Q_0$ (and thus may err in the overall production rate).  However,
 such partons do not branch any further, do not produce any soft
 offspring, so that the {\em shape}\/ of the resulting energy
 distribution remains undamaged.  
 We see that colour coherence plays here a crucial r\^ole.

 Thus, according to LPHD, the $x$-shape of the so-called ``limiting''
 parton spectrum which is obtained by formally setting $Q_0=\Lambda$
 in the evolution equations, should be mathematically similar to that
 of the inclusive distribution of (light) hadrons~$h$.  Another
 essential property is that the ``conversion coefficient'' $K^h$
 should be a true constant independent of the hardness of the process
 producing the jet under consideration.

\subsection{Another world}
  
It is important to realize that knowing the spectrum of {\em partons},
even knowing it to be a CIS quantity in certain sense, does not
guarantee on its own the predictability of the {\em hadron}\/
spectrum.  It is easy to imagine a world in which each quark and gluon
with energy $k$ produced at the small-distance stage of the process
would have dragged behind its personal ``string'' giving birth to $\ln
k$ hadrons in the final state (the Feynman plateau).
The hadron yield then would be given by a convolution of the parton
distribution with a logarithmic energy distribution of hadrons from
the parton fragmentation. 

If it were the case, each parton would have contributed to the yield
of non-relativistic hadrons and the hadron spectra would peak at much
smaller energies, $\xi_{\max}\simeq \ln Q$, in a spectacular
difference with experiment.

 Physically, it could be possible if the non-perturbative (\NP)
 hadronization physics did not respect the basic rule of the
 perturbative\1 dynamics, namely, that of colour coherence.

There is nothing wrong with the idea of convoluting time-like parton
production in jets with the inclusive \NP\ parton$\to$hadron
fragmentation function, the procedure which is similar to convoluting
space-like parton cascades with the \NP\ initial parton distributions
in a target proton to describe DIS structure functions.

What the nature is telling us, however, is that this \NP\ fragmentation
has a finite multiplicity and is {\em local}\/ in the momentum space.
Similar to its \PT\ counterpart, the \NP\ dynamics has a short memory: the
\NP\ conversion of partons into hadrons occurs locally in the
configuration space.

 The fact that even a legitimate finite smearing due to hadronization
 effects does not look mandatory makes one think of a deep duality
 between the hadron and quark-gluon languages applied to such a global
 characteristic of multi-hadron production as an inclusive energy
 spectrum.

 The message is, that ``brave gluon counting'', that is applying the
 \PT\1 language all the way down to very small transverse momentum
 scales, indeed reproduces the $x$- and $Q$-dependence of the observed
 inclusive energy spectra of charged hadrons (pions) in jets.

 Experimental evidence in favour of LPHD is mounting, and so is the
 list of challenging questions to be answered by the future
 quantitative theory of colour confinement.

\section{PROBING THE NON-PERTURBATIVE\1 DOMAIN WITH PERTURBATIVE\1 TOOLS}

\begin{flushright}
 \begin{minipage}{5in} {\em There is no heresy in it, and if not
 manifestly defined in Scripture, yet it is an opinion of good and
 wholesome use in the cours and actions of a man's life, and would
 serve as an hypothesis to solve many doubts whereof common philosophy
 affordeth no solution.} \end{minipage}

Sir Thomas Browne, ca 1635~\cite{STBrowne}
\end{flushright}

 Let us discuss the test case of the total cross section of $e^+e^-$
 annihilation into hadrons as an example.

 To predict $\sigma^{\mbox{\scriptsize tot}}_{\mbox{\scriptsize
 hadr}}$ one calculates instead the cross sections of quark and gluon
 production, $(e^+e^-\to q\bar{q})$ + $(e^+e^-\to q\bar{q}+g)$ + etc.,
 where quarks and gluons are being treated {\em perturbatively}\/\1 as
 real (unconfined, flying) objects. The {\em completeness}\/ argument
 provides an apology for such a brave substitution:
\begin{quote}
  Once instantaneously produced by the electromagnetic (electroweak)
  current, the quarks (and secondary gluons) have nowhere else to go
  but to convert, {\em with unit probability}, into hadrons in the end
  of the day.
\end{quote}
This {\em guess}\/ looks rather solid and sounds convincing, but
relies on two hidden assumptions:
\begin{enumerate}
\item 
  The allowed hadron states should be numerous as to provide the
  quark-gluon system the means for ``regrouping'', ``blanching'',
  ``fitting'' into hadrons.
\item
  It implies that the ``production'' and ``hadronization'' stages of
  the process can be separated and treated independently.
\end{enumerate}
1. To comply with the first assumption the annihilation energy has to
be taken large enough, $s\equiv Q^2\gg s_0$. In particular, it fails
miserably in the resonance region $Q^2\la s_0\sim 2M_{\rm res}^2$.
Thus, the point-by-point correspondence between hadron and quark cross
sections,
$$
 \sigma^{\mbox{\scriptsize tot}}_{\mbox{\scriptsize hadr}}(Q^2) 
 \>\stackrel{\mbox{?}}{=}\> 
 \sigma^{\mbox{\scriptsize tot}}_{{q\bar{q}+X}}(Q^2), 
$$
cannot be sustained except at very high energies.  It can be traded,
however, for something
 more
 manageable.

 Invoking the dispersion relation for the photon propagator (causality
 $\Longrightarrow$ analyticity) one can relate the {\em energy
 integrals}\/ of $\sigmatot(s)$ with the correlator of electromagnetic
 currents in a deeply Euclidean region of large {\em negative}\/
 $Q^2$. The latter corresponds to small space-like distances between
 interaction points, where the perturbative\1 approach is definitely
 valid.

 Expanding the answer in a formal series of local operators, one
 arrives at the structure in which the corrections to the trivial unit
 operator generate the usual perturbative~\1 series in powers of
 $\alpha_s$ (logarithmic corrections), whereas the vacuum expectation
 values of dimension-full (Lorentz- and colour-invariant) QCD
 operators provide non-perturbative~\1 corrections suppressed as
 powers of $Q$.

 This is the realm of the famous ITEP sum rules~\cite{ITEP} which
 proved to be successful in linking the parameters of the low-lying
 resonances in the Minkowsky space with expectation values
 characterising a non-trivial structure of the QCD vacuum in the
 Euclidean space.
 The leaders among them are the gluon condensate $\alpha_s
 G^{\mu\nu}G_{\mu\nu}$ and the quark condensate
 $\lrang{\psi\bar\psi}\lrang{\psi\bar\psi}$ which contribute to the
 total annihilation cross section, symbolically, as
\begin{equation}
\label{eq:itep}
 \sigma^{\mbox{\scriptsize tot}}_{\mbox{\scriptsize hadr}}(Q^2) -
 \sigma^{\mbox{\scriptsize tot}}_{{q\bar{q}+X}}(Q^2) \>=\>
 c_1{\frac{\alpha_s G^2}{Q^4} + c_2\frac{\lrang{\psi\bar\psi}^2}{Q^6}}
 + \ldots\,.
\end{equation}

\noindent
 2. Validating the second assumption also calls for large $Q^2$.  To
 be able to separate the two stages of the process, it is {\em
 necessary}\/ to have the production time of the quark pair $\tau\sim
 Q^{-1}$ to be much smaller than the time $t_1\sim \mu^{-1}\sim
 1\,{\rm fm}/c$ when the first hadron appears in the system. Whether
 this condition is {\em sufficient}, is another valid question. And a
 tricky one.

 Strictly speaking, due to gluon bremsstrahlung off the primary
 quarks, the perturbative production of secondary gluons and
 $q\bar{q}$ pairs spans an immense interval of time, ranging from a
 very short time,
 $\tform\sim Q^{-1}\ll t_1$, 
all the way up to a macroscopically large time 
$\tform\la Q/\mu^2\gg t_1$.

 This accompanying radiation is responsible for formation of hadron
 jets. It does not, however, affect the total cross section.  It is
 the rare hard gluons with large energies and transverse momenta,
 $\omega\sim k_\perp\sim{Q}$, that only matter.  This follows from the
 celebrated Bloch-Nordsieck theorem
 which states that the logarithmically enhanced (divergent)
 contributions due to real production of {\em collinear}\/
 ($k_\perp\ll Q$) and {\em soft}\/ ($\omega\ll Q$) quanta cancel
 against the corresponding virtual corrections:
$$
 \sigma^{\mbox{\scriptsize tot}}_{q\bar{q}+X} = \sigma_{Born}
\left(1+\frac{\alpha_s}{\pi}\left[\infty_{\mbox{\scriptsize real}}
-\infty_{\mbox{\scriptsize virtual}} \right]+\ldots \right) =
\sigma_{Born}
\left(1+\frac{3}{4}\frac{C_F\alpha_s(Q^2)}{\pi}+\ldots\right).
$$
 The nature of the argument is purely perturbative\1. Can the
 Bloch-Nordsieck result hold beyond pQCD?

 Looking into this problem produced an extremely interesting result
 that has laid a foundation for the development of perturbative\2
 techniques aimed at analysing non-perturbative\1 effects.

 V.~Braun, M.~Beneke and V.~Zakharov have demonstrated that the
 real-virtual cancellation actually proceeds much deeper than was
 originally expected~\cite{BBZ}.

 Let me briefly sketch the idea.  
\begin{itemize}
\item
 First one introduces an infrared cutoff (non-zero gluon mass $m$)
 into the calculation of the radiative correction.
\item
 Then, one studies the dependence of the answer on $m$.  A CIS
 quantity, by definition, remains finite in the limit $m\to0$.  This
 does not mean, however, that it is insensitive to the modification of
 gluon propagation. In fact, the $m$-dependence provides a handle for
 analysing the {\em small transverse momenta}\/ inside Feynman
 integrals. It is this region of integration over parton momenta where
 the QCD coupling gets out of perturbative\1 control and the genuine
 non-perturbative physics comes onto the stage.
\item
 Infrared sensitivity of a given CIS observable is determined then by
 the first non-vanishing term which is {\em non-analytic}\/ in $m^2$
 at $m= 0$.
\end{itemize}
 In the case of one-loop analysis of $\sigmatot$ that we are
 discussing, one finds that in the sum of real and virtual
 contributions not only the terms singular as $m \to 0$,
$$
  \ln^2m^2\;,  \;\;\;\;\;\; \ln m^2
\,,
$$
 cancel, as required by the Bloch-Nordsieck theorem, but that the
 cancellation extends~\cite{BBZ,BB} also to the whole tower of {\em
 finite}\/ terms
$$
 m^2\ln^2 m^2\;, \;\;\;\; m^2\ln m^2\,, \;\;\;\; m^2\;, 
 \;\;\;\; m^4\ln^2m^2\;, \;\;\;\; m^4\ln m^2 \,. 
$$ 
 In our case the first {\em non-analytic}\/ term appears at the level
 of $m^6$:
$$ 
 \frac{3}{4}\frac{C_F\alpha_s}{\pi}\left( 1 + 2\frac{m^6}{Q^6}\ln
 \frac{m^2}{Q^2} +\cO{m^8}\right).
$$
 It signals the presence of the non-perturbative\1 $Q^{-6}$ correction
 to $\sigmatot$, which is equivalent to that of the ITEP quark
 condensate in~\eqref{eq:itep}. (The gluon condensate contribution
 emerges in the next order in $\alpha_s$.)

 A similar program can be carried out for other CIS quantities as
 well, including intrinsically Minkowskian observables which address
 the properties of the final state systems and, unlike the total cross
 sections, do not have a Euclidean image.

\subsection{Event shapes in $e^+e^-$ annihilation}

 The most spectacular non-perturbative\1 results were obtained for a
 broad class of
 {\em event shape variables}\/ (like Thrust $T$, $C$-parameter,
 Broadening $B$, and alike).  As has long been expected~\cite{hadro},
 these observables possess relatively large $1/Q$ confinement
 correction effects.

 Employing the ``gluon mass'' as a large-distance trigger was
 formalised by the so-called dispersive method~\cite{DMW}.
 There it was also suggested to relate new non-perturbative\1
 dimensional parameters with the momentum integrals of the effective
 QCD coupling $\alpha_s$ in the infrared domain. Though it remains
 unclear how such a coupling can be rigorously defined from the first
 principles, the {\em universality}\/ of the coupling makes this guess
 verifiable and therefore legitimate. All the observables belonging to
 the same class $1/Q^{p}$ with respect to the nature of the leading
 non-perturbative\1 behaviour, should be described by the same
 parameter.

 \paragraph{{\em Whose}\/ coupling is it?} Approaching the borderline
 where \PT\1 gluons are about to disappear, one may talk about {\em
 gluers}\/ as carriers of the \NP\1 \PT\2 colour field. A formal
 definition of gluers is as follows.

 A {\em gluer}\/ is a miserable {\em gluon}\/ which hasn't got enough
 time to truly behave like one because its hadronization time is
 comparable with its formation time, $ t_{\mbox{\scriptsize
 form.}}\simeq {\omega}/{k_\perp^2} \> {\sim}\> t_{\mbox{\scriptsize
 hadr.}}\simeq \omega R_{\mbox{\scriptsize conf.}}^2$.  Contrary to
 respectful \PT\ gluons with small transverse size, $k_\perp\gg
 R_{\mbox{\scriptsize conf.}}^{-1}$, gluers are not ``partons'': they
 do not participate in \PT\2 cascading (don't multiply). According to
 the above definition, gluers have {\em finite transverse momenta}\/
 (though may have {\em arbitrarily large energies}\/).

 The r\^ole of gluers consists in providing comfortable conditions for
 {\em blanching}\/ colour parton ensembles (jets) produced in hard
 interactions. By examining the space-time picture of the parton
 formation~\cite{LPHD} one can convince oneself that formation of a
 gluer is a signal of hadronization process taking place in a given
 space-time region, {\em locally}\/ in the configuration space (recall
 the problem of soft confinement!)

 Having transverse momenta of the order of the inverse confinement
 scale makes their interaction strength potentially large,
 $\as(R_{\mbox{\scriptsize conf.}}^{-1})\sim 1$.  A uniform
 distribution in (pseudo)rapidity, together with finite transverse
 momenta with respect to the direction of the charge (jet, subjet)
 makes the gluers representatives of the Lund
 string~\cite{Lund_string}.

 The basic idea (see~\cite{DMW} and references therein) was to relate
 (uncalculable) \NP\1 corrections to (calculable) \PT\1 cross
 sections/observables through the intensity of {\em gluer}\/ emission
 -- {$\as$} in the infrared domain.
 In particular, an extended family of event shapes (not to forget
 energy-energy correlations~\cite{EEC}, out-of-plane transverse
 momentum flows~\cite{kout} etc.) can be said to measure the first
 moment of the perturbative\2 non-perturbative\1 coupling,
\begin{equation}
 \alpha_0 \>\equiv\> \frac{1}{\mu_I} \int_0^{\mu_I} dk\,
 \alpha_s(k^2), \qquad \mu_I= 2\,\mbox{GeV},
\end{equation}
 where the choice of the infrared boundary value $\mu_I$ is a matter
 of convention.


\paragraph{The Broadening story: a mistake but not an error.}

 A wonderful example of a mistake, in a sense of the introductory
 Section, was provided by the recent turbulent story of the Broadening
 measure.

 $B$ is defined as the sum of the moduli of transverse momenta of
 particles wrt the Thrust axis of the $e^+e^-$ annihilation event.
 Originally the \NP\ contribution to $B$ was naturally thought to
 accumulate gluers with rapidities up to $\log Q$. As a result
 theorists expected the \PT\1 distribution in $B$ to acquire a $\ln
 Q$-enhanced \NP\ shift.

 The data however refused to go along~\cite{H1,squeeze}. Fits based on
 the $\log Q$-enhanced shift were bad and produced too small a value
 of $\as(M_Z)$, and the \NP\ parameter $\alpha_0$ inconsistent with
 that extracted from analyses of the Thrust and $C$-parameter means
 and distributions.

 Universality of $\alpha_0$ and thus viability of the very notion of
 the infrared-finite coupling was seriously questioned.
 Pedro Movilla Fern\'andez who reported the findings of the
 resurrected JADE collaboration at the QCD-1998 conference in
 Montpellier did not stop at that~\cite{squeeze}. He came up with the
 study of ``what is wrong'' with the Broadening measure as such.  A
 comparison of MC-generated $B$ distributions at parton and hadron
 levels produced an unexpected result. While the $T$ and $C$ cases
 showed the expected shift patterns, the bump of the {\em hadronic}\/
 $B$ distribution turned out to be not only shifted but also
 \underline{squeezed} as compared with the {\em partonic}\/ one. This
 looked pretty anti-intuitive: how can one get a {\em narrower}\/
 distribution after a smearing due to hadron production?

 A solution came with recognition of the fact that the $B$ measure is
 more sensitive to quasi-collinear emissions than other event shapes,
 and is therefore strongly affected by an interplay between \PT\ and
 \NP\ radiation effects. With account of the omnipresent \PT\ gluon
 radiation, the {\em direction of the quark}\/ that forms the jet
 under consideration can no longer be equated with the direction of
 the Thrust axis (employed in the definition of $B$).  As a result,
 the range of the pseudorapidity of gluers contributing to the \NP\
 shift decreases from $\ln Q$ downto $\ln(1/B)$. The shift becomes
 $B$-dependent giving rise to a narrower distribution all
 right~\cite{broad}.

\vspace{2mm}
\noindent
\begin{minipage}{0.48\textwidth}
\epsfig{file=btdist-jade35.eps,width=1.1\textwidth,height=\textwidth,angle=90}
\end{minipage}
\hfill
\begin{minipage}{0.48\textwidth}
\epsfig{file=btdist-opal91.eps,width=1.1\textwidth,height=\textwidth,angle=90}
\end{minipage}\\[2mm]
Fig.12: Perturbative (dashed) and \NP-shifted/squeezed (solid) total
Broadening distributions~\cite{broad}.

Three lessons were drawn from the Broadening drama.
\begin{itemize}
\item 
  A pedagogical lesson the Broadenings taught, was that of the
  importance of keeping an eye on \PT\ gluons when discussing effects
  of \NP\ gluers.  An example of a powerful interplay between the two
  sectors was recently given by the study of the energy-energy
  correlation in $\epem$ in the back-to-back
  kinematics~\cite{EEC}. The leading $1/Q$ \NP\ contribution was shown
  to be promoted by \PT\1 radiation effects to a much slower falling
  correction, $Q^{-0.32\mbox{-}0.36}$.
 
\item 
  The physical output of the proper theoretical treatment was the
  restoration of the universality picture: within a reasonable 20\%\
  margin, the \NP\ parameters extracted from $T$, $C$ and $B$ means
  and distributions were found to be the same.

\item 
  A gnostic output was also encouraging. Phenomenology of \NP\
  contributions to event shapes has shown that it is a robust field
  with a high discriminative power: it does not allow one to be misled
  by theorists.
\end{itemize}
 Looking forward to the Conclusions Section please keep in mind the
 key words ``resurrected collaboration'' and ``error-free LEP data''.

\subsection{DIS jet shapes and non-global logs}

 Theoretical study of jet shapes in DIS was pioneered by Vito
 Antonelli, Mrinal Dasgupta and Gavin Salam by the 
 derivation of resummed \PT\ prediction for the Thrust distribution of
 the current fragmentation jet~\cite{Antonelli}. Two years later the
 Broadening measure followed~\cite{DISbroad}.

 On the way from $T$ to $B$, Dasgupta and Salam stumbled upon a new
 source of significant purely perturbative\1 next-to-leading (SL)
 corrections which was previously overlooked in the literature.  They
 dubbed these corrections ``non-global
 logs''~\cite{NonGlobal}\footnote{To put a long story short, the
 origin of these non-global effects has to do with the fact that in
 DIS one is forced to deal with characteristics of a \underline{single
 jet} rather than shapes of the hard \underline{event as a whole}. In
 other words, one resticts the measurements to a part of the available
 phase space; see~\cite{NonGlobal} for details.}.

  The final wisdom about DIS jet shapes can be found in~\cite{DISevsh}.

\subsection{On the universality of the infrared coupling -- 2003}

\begin{center}
\epsfig{file=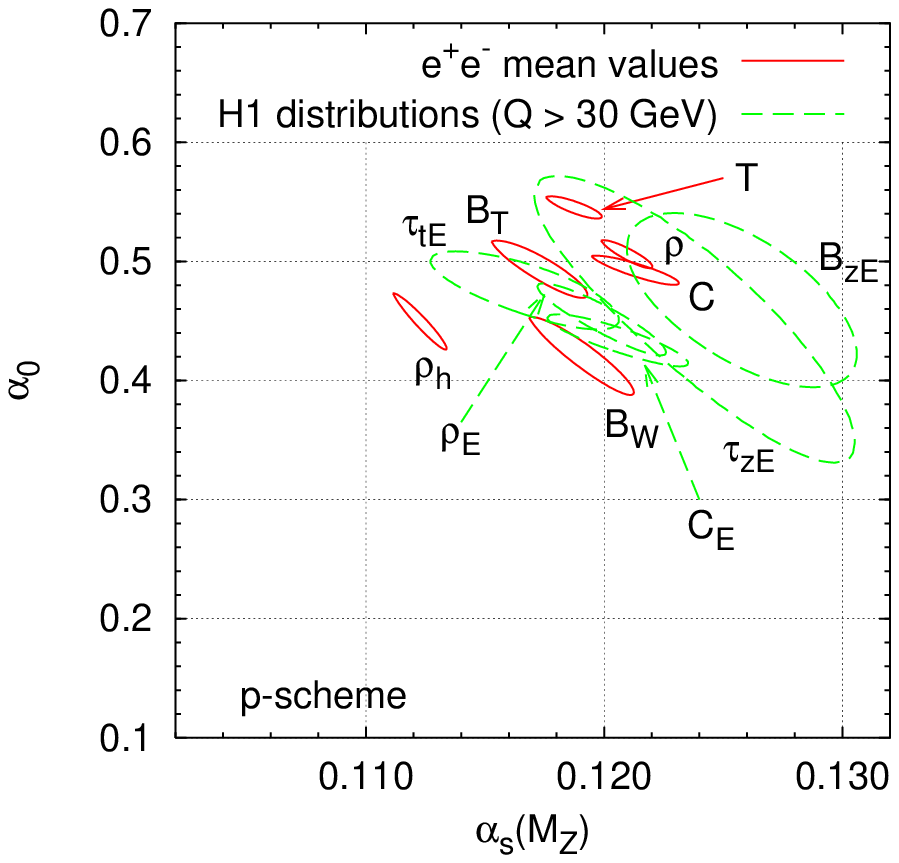,width=0.6\textwidth}

One standard deviation ellipses in the $\as(M_Z)$ -- $\alpha_0$
plane~\cite{GPS}.
\end{center}

 \section{Conclusions}

 Dr.\ John Dee, a British scholar, mathematician, an alchemist,
 hermeticist, cabalist and adept in esoteric and occult lore, was
 Queen Elizabeth's philosopher and astrologer.

\noindent
\begin{minipage}{0.65\textwidth}
\parindent 3mm
 A visionary
 of the Empire, he coined the word Brittannia, was the first to apply
 Euclidean geometry to navigation, trained the great navigators,
 developed a plan for the British Navy and established the legal
 foundation for colonizing North America. Dee wrote a famous
 Mathematical Preface to his translation of Euclid in which he
 systematised future development of the sciences based on mathematics.
 He also extensively practiced as an angel conjuror (with Edward
 Kelley for many years his {\em skryer}) and, some say~\cite{PaulLee},
 was the one who in 1588 ``{\em put a hex on the Spanish Armada which
 is why there was bad weather and England won}\/''\footnote{Speaking
 of practicality of communicating with angels: ``{\em {\em [John Dee]}
 also speaks of seeing the sea, covered with many ships. Uriel warns
 them {\em [Dee \& Kelley]} that foreign Powers are providing ships
 `against the welfare of England, which shall shortly be put in
 practice.'  \ldots The defeat of the Spanish Armada took place \ldots
 four years after this vision.}''~\cite{char1}}.
\end{minipage}
\hfill
\begin{minipage}{0.3\textwidth}
\includegraphics[width=\textwidth]{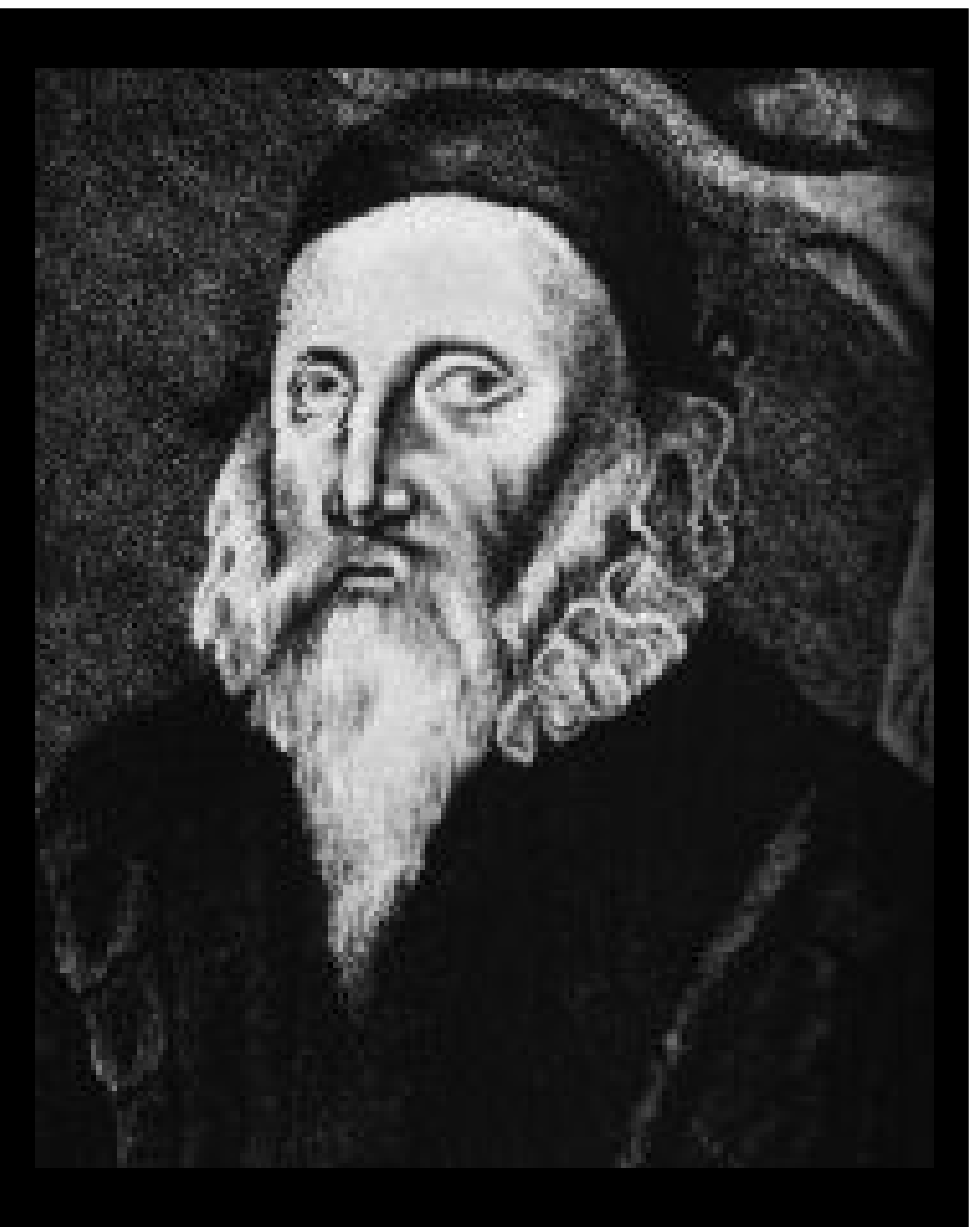}

\begin{center}
 John Dee (1527 -- 1608)
\end{center}
\end{minipage}

 The volume of this writeup prevents us from going deeper into the
 fascinating story of John Dee's life and endeavours. Dee's story is
 relevant to the present lectures: there is an important message to
 take home.

 The ``crystal egg''~\footnote{another source tells of ``an
 Aztec/Mayan obsidian show stone''} John Dee used to communicate with
 spirits (and the cherub he identified as Archangel Uriel, in
 particular) rests, reportedly, in the British Museum along with his
 conjuring table~\cite{PaulLee,Wilson}.  These were John Dee's {\em
 detector gadgets}\/ if you please. More importantly, the {\em
 experimental data}\/ that Dee collected in 1580's, his manuscripts
 and diaries are also being carefully preserved in the British
 Library~\cite{chm,hm}.

 One might question the value of Dee's {\em De Heptarchi\ae\
 Mystic\ae}\/ (i.e.\ Detailed instructions for communicating with
 angels and employing their aid for practical purposes) as a source of
 inspiration and knowledge acquisition for the future. What cannot be
 questioned, however, is that the {\em experimental data}\/ collected
 by LEP exactly 400 years after Dr.\ Dee was communicating with Uriel
 \& Co, will remain, for many a year to come, an unmatched source of
 knowledge about the physics of hadrons.

 Will there be a caring ``{\em British Library}\/'' to preserve LEP
 ``diaries'' for theorists who will sooner or later come close to
 deciphering the ``Enochian Alphabet'' of QCD confinement?

\begin{quote}
 {\em An angel tells him {\em [Dr.\ Dee]} they are to be ``rocks in
 faith.'' ``While sorrow be meansured thou shalt bind up thy
 fardell.'' He is not to seek to know the mysteries till the very hour
 he is called. ``{\bf Can you bow to Nature and not honour the
 workman?}''}~\cite{char1}
\end{quote}


\section*{ACKNOWLEDGEMENTS}

 I would like to thank Vladimir Dokchitser, Alexei Safonov and Gavin
 Salam for their help in the long course of preparing this writeup and
 to Nick Ellis for his patience in receiving it, as well as for the
 perfect organisation of the School. I am grateful to Jean Iliopoulos
 for Greek lessons and to Natalia Afanassieva for the enlightenment in
 regards of the r\^ole of Arch.\ Uriel in the development of
 fundamental science.

\end{document}